\documentclass[aps,showpacs,eqsecnum,twocolumn,superscriptaddress]{revtex4}

\usepackage{amsmath}
\usepackage{latexsym}
\usepackage{graphicx}
\usepackage{epstopdf}

\begin{document}


\title{Dynamical evolution of quasi-circular binary black hole data}


\author{Miguel Alcubierre}
\affiliation{Instituto de Ciencias Nucleares, Universidad
Nacional Aut\'onoma de M\'exico, A.P. 70-543, M\'exico D.F. 04510,
M\'exico}

\author{Bernd Br\"ugmann}
\affiliation{Theoretical Physics Institute, University of Jena, 
07743 Jena, Germany}

\author{Peter~Diener}
\affiliation{Center for Computation and Technology, 302 Johnston Hall,
Louisiana State University, Baton Rouge, LA 70803, USA}

\author{F.~Siddhartha~Guzm\'an}
\affiliation{Instituto de F\'{\i}sica y Matem\'{a}ticas, Universidad
Michoacana de San Nicol\'as de Hidalgo. Edificio C-3, Cd. Universitaria.
C. P. 58040 Morelia Michoac\'{a}n, M\'{e}xico.}

\author{Ian~Hawke}
\affiliation{Max-Planck-Institut f\"ur Gravitationsphysik,
Albert-Einstein-Institut, Am M\"uhlenberg 1, 14476 Golm, Germany}
\affiliation{School of Mathematics, University of Southampton,
Southampton SO17 1BJ, UK}

\author{Scott~Hawley}
\affiliation{Center for Relativity, University of Texas at Austin,
Austin, TX 78712, USA}

\author{Frank~Herrmann}
\affiliation{Max-Planck-Institut f\"ur Gravitationsphysik,
Albert-Einstein-Institut, Am M\"uhlenberg 1, 14476 Golm, Germany}

\author{Michael~Koppitz}
\affiliation{Laboratory for High Energy Astrophysics, NASA Goddard
Space Flight Center, 8800 Greenbelt Rd., Greenbelt, MD 20771, USA}

\author{Denis~Pollney}
\affiliation{Max-Planck-Institut f\"ur Gravitationsphysik,
Albert-Einstein-Institut, Am M\"uhlenberg 1, 14476 Golm, Germany}

\author{Edward~Seidel}
\affiliation{Center for Computation and Technology, 302 Johnston Hall,
Louisiana State University, Baton Rouge, LA 70803, USA}
\affiliation{Max-Planck-Institut f\"ur Gravitationsphysik,
Albert-Einstein-Institut, Am M\"uhlenberg 1, 14476 Golm, Germany}
\affiliation{Department of Physics and Astronomy
Louisiana State University, Baton Rouge, LA 70803, USA}

\author{Jonathan Thornburg}
\affiliation{Max-Planck-Institut f\"ur Gravitationsphysik,
Albert-Einstein-Institut, Am M\"uhlenberg 1, 14476 Golm, Germany}


\date{\today}


\begin{abstract}
We study the fully nonlinear dynamical evolution of binary black hole
data, whose orbital parameters are specified via the effective
potential method for determining quasi-circular orbits. The cases
studied range from the Cook-Baumgarte innermost stable circular orbit
(ISCO) to significantly beyond that separation. In all cases we find
the black holes to coalesce (as determined by the appearance of a
common apparent horizon) in less than half an orbital period. The
results of the numerical simulations indicate that the initial holes
are not actually in quasi-circular orbits, but that they are in fact
nearly plunging together. The dynamics of the final horizon are
studied to determine physical parameters of the final black hole, such
as its spin, mass, and oscillation frequency, revealing information
about the inspiral process.  We show that considerable resolution is
required to extract accurate physical information from the final black
hole formed in the merger process, and that the quasi-normal modes of
the final hole are strongly excited in the merger process.  For the
ISCO case, by comparing physical measurements of the final black hole
formed to the initial data, we estimate that less than 3\% of the
total energy is radiated in the merger process.
\end{abstract}


\pacs{
04.25.Dm, 
04.30.Db, 
04.70.Bw, 
95.30.Sf, 
97.60.Lf  
%
\\
Preprint number: IGPG-04/10-4, AEI-2004-115
}


\maketitle


\section{Introduction}
\label{sec:introduction}

One of the most pressing problems in gravitational physics is the
solution of the two body problem for an inspiralling pair of black
holes (BHs) from quasi-circular orbit to merger. This process is
expected to be a likely candidate for early detection by the current
generation of gravitational wave detectors. Although considerable
success has been made in long term, 3D evolutions of single distorted
holes or head-on collisions~\cite{Anninos96c,Alcubierre02a}, and
hybrid numerical-perturbative evolutions~\cite{Baker00b,Baker:2001nu},
long-term evolutions of binary BH systems with angular momentum have
seen only limited
successes~\cite{Bruegmann97,Brandt00,Alcubierre00b,Seidel99b}, up to
timescales of a single orbit~\cite{Bruegmann:2003aw}. With limited
evolution times, emphasis has been on starting from solutions to the
constraint equations which approximate the very last phase of the
inspiral process, close to the final plunge.

Such a family of BH binary parameters was proposed by
Cook~\cite{Cook94}, and is based on finding minima of an effective
potential for boosted binary BHs (with throats).  The effective
potential approach was also applied later in~\cite{Baumgarte00a} for
binaries constructed from puncture data, where similar results were
found for the innermost stable circular orbit, or ``ISCO''.
Furthermore, recent work~\cite{Tichy:2003qi,Tichy03a} finds a very similar
sequence of orbital parameters for binary puncture BHs using an
entirely different construction based on the assumption of a ``helical
Killing vector'' (HKV). On the other hand,
Refs.~\cite{Gourgoulhon02,Grandclement02,Cook:2001wi}, the first to
develop the HKV approach, solve the thin sandwich equations for binary
BHs (with throats), obtaining markedly different results for the
estimate of the ISCO.  These results agree with increasingly higher
order post-Newtonian calculations~\cite{Damour:2002qh}. At somewhat
larger separations, the various constructions show better agreement,
as would be expected.

It is critical to understand the differences between these data sets,
in particular to investigate whether any of them represent late stages
of physically realistic orbital BH mergers.  The only method to study
the fully nonlinear orbital dynamics of these data sets, which are
well into the strong-field region, is full numerical
simulation. Recent advances, including the use of the BSSN evolution
system~\cite{Nakamura87, Shibata95, Baumgarte99, Alcubierre99d}, new
gauge conditions~\cite{Alcubierre02a}, and excision
techniques~\cite{Alcubierre01a,Alcubierre2003:BBH0-excision}, have
brought about significant improvements in the ability of numerical
relativity to handle binary systems. Further, by using co-rotating
coordinates (implemented through an adjustment to the shift
condition~\cite{Alcubierre2003:co-rotating-shift}) we find that the
BHs can effectively be held in place, greatly reducing dynamics in
grid functions and removing complications due to the need of moving
excision regions.  These improvements have allowed us to carry out the
first systematic and fully nonlinear numerical studies of the orbital
dynamics of such data sets, constructed to be in the
``near-ISCO-regime''.  The dynamics of the entire inspiral and merger
process can be followed to well over $t \approx 100 M$ in many cases.
(Distances and times are given in terms of the ADM mass, which is
approximately unity for the initial data models studied here.)
Similar techniques were recently used to evolve a pair of separated
apparent horizons for the expected period of a close
orbit~\cite{Bruegmann:2003aw}.

In this paper we apply these full nonlinear capabilities to examine
the orbital dynamics of a near-ISCO sequence. The sequence, based on
the effective potential work of Refs.~\cite{Cook94, Baumgarte00a}, is
constructed to represent a family of circular orbits of non-rotating,
equal mass puncture BHs. These parameters were previously
studied~\cite{Baker:2001nu, Baker:2002qf} with a hybrid
numerical-perturbative approach (see,
e.g.,~\cite{Abrahams92a,Abrahams95d,Baker00b}).  Increasingly
sophisticated over the years, this hybrid approach is aimed at
extracting waveforms if a linear regime can be reached, while only
fully relativistic approaches are capable of studying the dynamics of
the complete inspiral process.  The near-ISCO regime studied here is
for initial proper spatial separations ranging from $4.99M$ (the
Cook-Baumgarte ISCO) to $7.84M$. The data are evolved at least until a
common horizon forms, indicating that the pair has merged. In all
cases we find that the merger occurs in roughly half (or less) of the
expected period, based on the determination of the angular velocity in
the initial data calculation.  These and other measures discussed
below indicate that rather than orbiting BHs, these data sets
represent nearly plunging BHs. 

Using techniques originally developed
in~\cite{Anninos94f,Anninos95c,Brandt94c,Libson94a,Masso95a,Bruegmann96}
and applied to grazing BH collisions
in~\cite{Alcubierre00b,Bruegmann97,Seidel99b,Bruegmann99b}, we study
the physical properties, such as mass and spin, and the dynamics of
the final BH horizons formed in the merger process.  In particular, we
compare a number of independent measures of horizon dynamics,
including the geometry of apparent horizons and event horizons, and
the isolated horizon
formalism~\cite{Ashtekar98a,Dreyer-etal-2002-isolated-horizons,Ashtekar:2004cn},
which show consistent results for masses and spins of the final black
holes.  Comparing these results with the parameters of the initial
data, we are able to infer some properties of the orbital dynamics.
We find that at low resolution, the angular momentum of the final BH
is significantly lower than the angular momentum of the spacetime.
However, for the case studied in most detail at higher resolution, the
angular momentum of the final BH formed in the merger is within about 20\%
of the angular momentum of the spacetime.  Hence, with current
resolutions, we are not yet able to use this to measure the angular
momentum radiated.  On the other hand, our highest resolution
calculations of the final BH mass indicate that less than 3\% of
$M_\text{ADM}$ is radiated, showing very good accuracy of the overall
energy accounting in the simulations, but the current resolutions are
still not sufficient to use this measure to predict precisely the
radiated energy.  By studying the dynamics of the horizon geometry, we
are further able to determine the oscillation frequency of the final
BH, showing that its quasi-normal modes (QNMs) are strongly excited in
the merger process.


\section{Initial Data}
\label{sec:inital_data}

The initial data are calculated via the puncture
method~\cite{Brandt97b}. This variant of Brill-Lindquist initial data
performs a conformal compactification at the interior asymptotically
flat region of the BHs. The extrinsic curvature is given by the
Bowen-York solution to the momentum constraint, and the Hamiltonian
constraint is solved for the conformal factor. The data are
conformally flat and maximally sliced. The initial lapse is chosen to
be $\alpha=1$, while the initial shift is determined by a quasi-rigid
rotation (see below).

Each member of the initial data sequence we consider has two BHs of
equal mass and equal opposite momentum, without spin. Initial
parameters for such systems in quasi-circular orbit are tabulated
in~\cite{Baker:2002qf}, based on the effective potential calculations
of~\cite{Cook94} and~\cite{Baumgarte00a} for punctures.  In
Table~\ref{tbl:QCtable} we list the first five elements in this
sequence.  The effective potential construction singles out QC-0 as
representing the presumed innermost stable circular orbit, or ISCO,
while other members of the sequence of quasi-circular orbits have
initial radii progressively further out.  We evolve all members of
this near-ISCO regime to coalescence and beyond, using full numerical
relativity.  The measure $L/M$ between the holes is the proper
distance computed along a coordinate line from the location of one
apparent horizon (AH) to the other.  For convenience, we also provide
dynamically determined coalescence times in the Table, as measured by
the appearance of the horizons, both as computed by coordinate time
and by the proper time measured by an observer at the origin between
the two BHs.  Further discussion of the evolution results are the
subject of the following sections, but one can see immediately that by
these measures, coalescence occurs well before the orbital time scale
for all cases studied.

\begin{table}
  \begin{ruledtabular}
  \begin{tabular}{cccccccc}
  Model & $L/M$  & $J/M^2$ & $M\Omega$ &  $T_\text{orbit}$ & $t_\text{AH}$
        & $\tau_\text{AH}^\text{cent}$ & $\tau_\text{EH}^\text{cent}$  \\
  \hline 
  QC-0  & 4.99   &  0.779  &  0.168 & 37.4 &  16.8 & \phantom{0}6.71 & 4.91 \\
  QC-1  & 5.49   &  0.781  &  0.142 & 44.2 &  19.2 & \phantom{0}8.49 & 6.66 \\
  QC-2  & 5.86   &  0.784  &  0.127 & 49.5 &  21.1 & \phantom{0}9.76 & -- \\
  QC-3  & 6.67   &  0.794  &  0.102 & 61.6 &  25.4 & 12.9\phantom{0} & -- \\
  QC-4  & 7.84   &  0.817  &  0.076 & 82.7 &  31.7 & 17.3\phantom{0} & -- \\
  \end{tabular}
  \end{ruledtabular}
  \caption{Elements of the initial data sequence
    of~\cite{Baker:2002qf}, listing the initial proper separations,
    angular momenta, and angular velocity ($L/M$, $J/M$, $M\Omega$,
    and corresponding orbital timescale $T_\text{orbit}$).  The
    coordinate time $t_\text{AH}$ and the central-observer proper time
    $\tau_\text{AH}^\text{cent}$ to common apparent horizon formation
    are given for the intermediate resolution $dx=0.06M$ runs with
    shift parameters $(\eta, p, n)=(4,1,4)$, corresponding to
    Figs.~\ref{fig:merger_times_orbit}
    and~\ref{fig:merger_times_others}. Also shown is the proper time
    $\tau_\text{EH}^\text{cent}$
    until a common event horizon is found for a central observer.
    Due to computational limitations,
    we did not attempt to locate EHs for QC-2 - QC-4.}
  \label{tbl:QCtable}
\end{table}


\section{Evolution methods}
\label{Sec:Evolution}

\emph{Formulation and gauge.} The evolution of the system is carried
out using the techniques and evolution system (``BSSN'')
\cite{Nakamura87, Shibata95, Baumgarte99} with the implementation
described explicitly in~\cite{Alcubierre02a}. We use a hyperbolic
version of the ``1+log'' lapse and $\Gamma$-driver shifts.  In
particular, the lapse $\alpha$ is evolved using
\begin{equation}
 \partial_t \alpha = - 2\alpha K,
\end{equation}
where $K$ is the trace of the extrinsic curvature. The shift is
implemented using
\begin{equation}
\partial_t \beta^i = F B^i \qquad
\partial_t B^i     = \partial_t \tilde{\Gamma}^i - \eta B^i \label{eq:shift} 
\end{equation}
with $\eta$ a constant dissipative parameter and $F$ a function of
space and time given by
\begin{equation}
  F = \frac{3}{4} \frac{\alpha^p}{\psi_{\text{BL}}^n}, \label{eq:F_shift}
\end{equation}
where $\psi_{\text{BL}}$ is the time independent conformal factor for
Brill-Lindquist data. This choice of $F$ ensures a shift which is
nearly static at late times, as well as ensuring the correct fall-off
near the puncture. The factor $3/4$ guarantees that the asymptotic
gauge speed associated with the longitudinal shift components is equal
to the speed of light.  These features were found to be important for
long-term stable and accurate evolutions of head-on collisions without
excision~\cite{Alcubierre02a}. Typically, we use values of~$p=1$~or $2$ and
$n=2$~or $4$. The parameter $\eta$ can be used to tune the rate of horizon
expansion over the course of the evolution, with larger values leading
to faster horizon growth. We typically use values $\eta\in[2,5]$.

As detailed in~\cite{Alcubierre2003:co-rotating-shift}, we 
apply a quasi-rigid rotation to the initial values of the shift
vector,
\begin{equation}
  \label{eq:shift_rotation}
  \left.\beta^x\right|_{t=0} 
    = - y \frac{1}{\psi_{\text{BL}}} \hat{\Omega} \qquad
  \left.\beta^y\right|_{t=0} 
    = x \frac{1}{\psi_{\text{BL}}} \hat{\Omega}.
\end{equation}
The factor of $1/\psi_{\text{BL}}$ modifying the rigid rotation
angular velocity $\hat{\Omega}$ has the effect of ensuring that the
shift is zero at the location of the punctures (a condition that is
necessary if one wishes that the punctures remain at a fixed location
in coordinate space). We find that with appropriate values of
$\hat{\Omega}$ in the initial values of the shift, the location of the
horizons can be made to remain almost stationary on the grid. Since
the size and location of the excision region is determined by the
apparent horizon location, this allows the evolution to proceed until
merger without requiring the excision region to move on the grid.
This type of co-rotating shift, with dynamic control to adjust the
co-rotating frame as the binaries orbit, is described
in~\cite{Alcubierre2003:co-rotating-shift}. To initialize the
co-rotation $\hat\Omega$, experiments were carried out to empirically
determine values for which the horizon tended to move least.

It is worth noting that the optimal $\hat\Omega$ values tended to be
$50-60\%$ of the values of $\Omega$ specified in
Table~\ref{tbl:QCtable}.  As the latter are asymptotically determined,
and further as slicing and shift conditions in the interior have
strong additional influences on the evolution of the BH orbit (for
example, the $\Gamma$-driver shift condition used in the evolutions
has itself a co-rotating
component~\cite{Alcubierre2003:co-rotating-shift}), it is difficult to
draw firm conclusions from this result.  However, the result may be
consistent with one of the main results of this paper, namely that the
near-ISCO regime orbital sequences studied here coalesce rapidly, well
within an orbital time period, as discussed in more detail below.

Apparent horizons are determined at each time-step using the horizon finder
{\tt AHFinderDirect} described in~\cite{Thornburg2003:AH-finding}.
As a consistency check,
apparent horizons are also determined each $0.48M$ using the earlier,
accurate but much less efficient horizon finder \texttt{AHFinder},
an implementation of the method described in~\cite{Gundlach97a,Alcubierre98b}.
For QC-0 and QC-1, an event horizon finder~\cite{Diener03a} was used
for further analysis, as a post-processing step.

\begin{figure}[t]
  \includegraphics*[width=18pc,height=14pc]{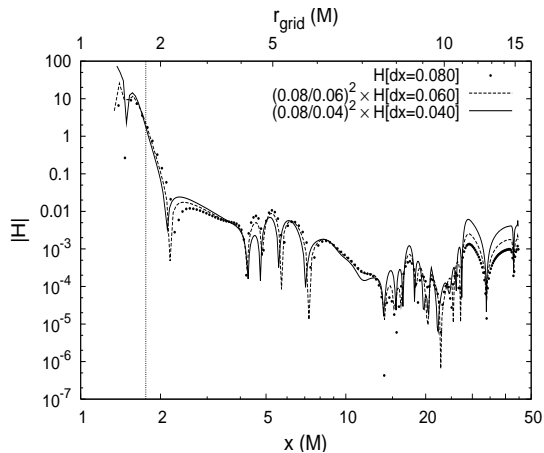}
  \caption{Convergence in Hamiltonian constraint for the QC-0 case
  shown at the time of formation of the common AH, $t=17.76M$.  The
  resolutions are $dx=0.08M, 0.06M$ and $0.04M$, with the evaluated
  constraint scaled appropriately so that the lines would lie on top
  of each other in the case of perfect second order convergence.  The
  lower scale shows the physical distance along the $x$ axis, while
  the upper scale shows the grid coordinate $r_\text{grid}$ defined
  by~\eqref{eq:fisheye}.  The vertical dotted line shows the apparent
  horizon position.  Note that although we do not obtain second order
  convergence near the outer boundary, there is good second order
  convergence in the neighborhood of the horizons.
  \label{fig:ham_convergence}}
\end{figure}

\emph{Numerical methods.} Finite differencing is performed to second
order accuracy, and the time stepping is carried out via a 3-step
iterative Crank-Nicholson scheme.  The boundaries
are placed at the same coordinate location for all simulations reported here. The
proper distance to the outer boundary is extended by applying a
``fisheye'' coordinate transformation to the initial
data~\cite{Baker00b, Alcubierre02a}. At the outer boundaries, an
outgoing-radiation (Sommerfeld) boundary condition is applied to each
of the evolution variables. The boundary condition introduces a
constraint violation, visible as a region where the constraints do not
converge to zero, which propagates inwards from the outer boundary. To
minimize the effect of this error on the interior dynamics, we have
placed the outer boundaries of our grid at a distance which is
causally disconnected (with respect to the physical speed of light)
from the common apparent horizon at the time of merger. Although this
does not guarantee complete numerical causal disconnection, it
significantly reduces potential outer boundary effects in the
neighborhood of the horizons during the pre-merger phase, and allows
us to achieve second order convergence (for instance in the
Hamiltonian constraint violation) in the region near the final merged
horizon (Fig.~\ref{fig:ham_convergence}).  Through convergence studies
and experiments with boundaries further away, we conclude that with
our computational grid parameters, the outer boundary does not affect
the results presented here.

\emph{Excision.} For the long-term simulations carried out in this
paper, we excise the singularity using an extension of the ``simple
excision'' techniques which have proven successful in evolving single
BH
spacetimes~\cite{Alcubierre00a,Alcubierre01a,Alcubierre2003:BBH0-excision}.
We use the same condition on each of the fields. Rather than excise a
cube we excise within a topologically spherical surface (a ``lego''
sphere) determined by the shape of the AH. A number of buffer points
(at least 5) is maintained between the AH and the excision boundary in
order to minimize the propagation of errors introduced by the excision
boundary condition out of the horizon. As shown
in~\cite{Alcubierre2003:BBH0-excision}, for the gauge conditions used
here this buffer zone is expected to be adequate to protect the BH
exterior from errors coming from the excision boundary.  The excision
boundary is allowed to grow as the AH grows, but not to move or shrink in coordinate
space -- that is, points which are excised are never returned to the
active grid.  We find that our runs are second order convergent even
in areas neighboring this excision boundary. Eventual loss of
convergence at late times, after merger, is entirely due to outer
boundary effects.


\section{Simulation domain}
\label{sec:setup}

For each dataset a number of test runs on a small grid and with low
resolution were performed in order to determine the approximate time
of common horizon formation. These times were used to construct grids
with outer boundaries causally disconnected from the common horizon at
merger time (distance to the outer boundary measured by the physical
distance~\footnote{Note that in our setup, gauge effects can (and do)
travel faster than light, so that influences from the boundary can in
fact affect the interior evolution at earlier times than the causal
separation might suggest. However, empirically we find that constraint
violations propagate at the speed of light, as can be seen from, e.g.,
Fig.~\ref{fig:ham_convergence}. Therefore the appearance of the common
horizon will not be affected by the boundary.}).  Initial datasets
were set up with punctures along the $y$-axis and momenta in the
$x$-direction. The $z$-reflection symmetry allowed the evolutions to
be carried out on the positive $z$ portion of the grid, with a
symmetry along the $z=0$ plane implemented through the use of
ghost-zones whose data where reflected from interior grid points.

For each model, runs were performed with uniform grid resolutions of
$dx = 0.08 M$ and $dx = 0.06 M$ on grids of $384\times 384 \times 192$
and $512\times 512 \times 256$ points respectively.  For 3-level
convergence tests, certain models were also run on $640\times 640
\times 320$ ($dx=0.048 M$) and $768\times 768 \times 384$ ($dx=0.040
M$), though especially the latter are at the limit of what even
large-scale clusters can sustain.

To allow the outer boundary to be farther from the strong-field region,
the (Cartesian) grid was placed non-uniformly in the physical coordinates:
the physical radius $r$ of each grid point was related to the grid radius
$r_\text{grid}$ by a ``transition fish-eye transformation''
(\cite{Alcubierre02a})
\begin{eqnarray}
r & = &
        a r_\text{grid}
        + (1 - a) \frac{s}{2 \tanh (r_{0}/s)}                   \nonumber \\
  &   & \quad
        \times
        \left[ 
        \ln \left( \cosh\frac{r_\text{grid}+r_{0}}{s} \right)
        -
        \ln \left( \cosh\frac{r_\text{grid}-r_{0}}{s} \right)
        \right]                                                 \nonumber \\
                                                        \label{eq:fisheye}
\end{eqnarray}
where the parameters $r_{0}=5.5M$ and $s=0.8M$ are the radius and
width of the transition region, and $a=4$ gives the physical scale
factor outside the transition region.  This placed the grid boundaries
$r_\text{grid}=15.48M$ at a physical distance $r=45.42 M$ from the
origin.

Fig.~\ref{fig:ham_convergence} displays a cross section of the
Hamiltonian constraint for QC-0 at $t=17.76M$ for
the three resolutions $dx=0.08M,0.06M,0.04M$, scaled to lie on top of
each other if the code is converging at second order. At this time the
common horizon has just appeared at the highest resolution. As is typical
for these runs for which the outer boundary condition is not
constraint preserving, a non-convergent effect marches inward from the
outer boundary. The region disconnected from the outer boundary,
however, shows second order convergence, including points in the
immediate neighborhood of the excision region. Significantly,
convergence is maintained in the region of common horizon
formation at the time at which it first forms.


\section{Results}
\label{sec:results}

Evolutions to merger were carried out for each of the QC-0 to QC-4
initial data models, at the $384$ and $512$ grid sizes described above
(a complete set of $640$ and $768$ runs was not possible due to
computational resource limitations).  We study the orbital dynamics
and merger times for all five members of this sequence.  Evolutions
were typically carried out to just beyond the formation of a common
apparent horizon (AH), though it is possible with careful parameter
choices to carry out the evolutions for significantly longer, often to
beyond $t=100M$.  At late times, however, contamination from the outer
boundaries can become a significant source of error.

\begin{figure}
  \includegraphics[width=18pc,height=12pc]{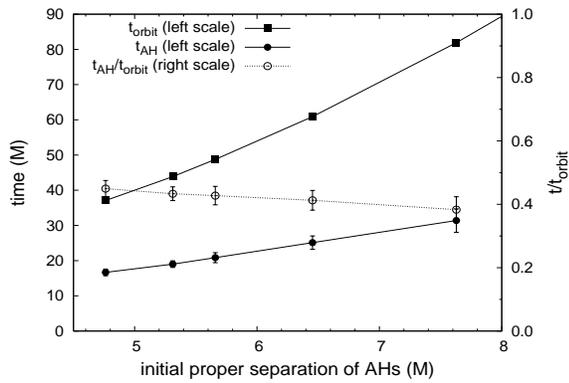}
  \caption{The time to appearance of a common AH for each of the QC
  models. Filled circles indicate the results of numerical experiments
  using $dx=0.06$, from initial proper spatial separations indicated
  along the $x$-axis. The upper line indicates the expected orbital
  period, based on the initial angular velocity (Table
  \ref{tbl:QCtable}) and assuming a Newtonian circular orbit.  Empty
  circles indicate the fraction of an orbit before common AH
  formation. The error bars show $\pm$~the difference between $dx=0.06$
  and $dx=0.08$ evolutions; the effects of variations in gauge parameters
  and outer-boundary position are generally smaller than these.}
  \label{fig:merger_times_orbit}
\end{figure}

\emph{Merger Times.}  The actual time at which the common AH forms for
each model is plotted in Fig.~\ref{fig:merger_times_orbit} for the
resolution $dx=0.06 M$.  The coordinate time at which a common AH
forms is clearly not a physical invariant, and depends not only on
gauge effects, but also on computational parameters.  For each point
in the near ISCO QC sequence we have therefore carried out a large
number of simulations, varying not only grid resolution and boundary
location, but also gauge conditions, within the class of conditions
discussed above.  We find that the results are only weakly dependent
on these effects; based on these tests we have provided approximate
error bars in Fig.~\ref{fig:merger_times_orbit} that span the general
range of results we obtained over all parameter variations.
Furthermore, we experimented with 4th order spatial and 3rd order
temporal differencing, which improves the accuracy of the lower
resolution results, but which does not change the essential results
presented here.  The variations in timing, over all gauges and
computational parameters tested, are typically less than 10\%, a
result that was typical of earlier studies~\cite{Anninos94b,
Alcubierre00b,Seidel99b} where, for example, rather different lapse
conditions were also tested with minimal changes in coalescence time
results.

For reference, the expected orbital periods based on the initial data
estimate of the angular velocity from Table~\ref{tbl:QCtable} are
plotted as square points (the top-most curve). For each of the models
studied, a common AH appeared in less than half of the expected time
for an orbit. This is true even for the further separated cases, well
beyond the effective potential ISCO, where dynamics would be expected
to better approximate a quasi-circular orbit.

\begin{figure}
  \includegraphics[width=18pc,height=12pc]{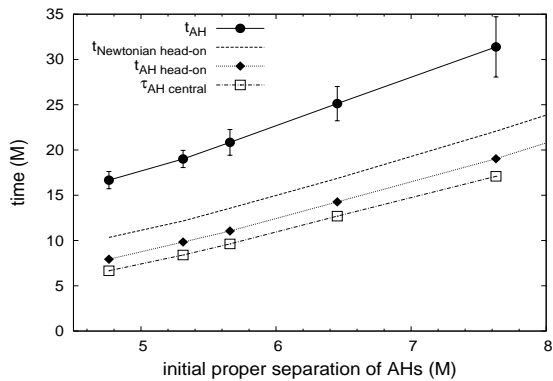}
  \caption{The time to common AH formation is compared with various
    head-on collision results. The upper line (filled circles) repeats
    the curve of Fig.~\ref{fig:merger_times_orbit}, showing common AH
    formation times for each of the QC models. The dashed line
    indicates the Newtonian collision time for a pair of particles
    falling together from the given distance with zero initial
    velocity. The corresponding relativistic head-on collisions were
    simulated and give the results plotted as filled diamonds. Finally,
    the lower curve (open boxes) refers to the same QC simulations as
    the upper (filled circle) curve, but uses the proper time at the
    origin at the time of common AH appearance (see text).}
  \label{fig:merger_times_others}
\end{figure}

To gain insight into the physical interpretation of the merger times,
the values obtained from full 3D evolutions of this near-ISCO sequence
can be compared to the merger times for head-on collisions, with each
puncture starting with zero linear momentum.  In Ref.~\cite{Anninos94b}
it was shown that there is a remarkable agreement between the
Newtonian free fall time, and the fully relativistic computation of
the time it takes for a common AH to form for BHs colliding head-on.
To do this comparison we carried out a set of 3D simulations, evolving
initial data for head-on collisions of BHs represented by punctures
initially at the same distances of the QC sequence studied in
Table~\ref{tbl:QCtable}, but without boosts. We used the same lapse
condition as described above for evolving the orbital configurations, 
while the shift condition was slightly modified~\footnote{
For historical reasons, instead of
equation~(\ref{eq:shift}), we have used
$$
\partial_t \beta^i = B^i \qquad
\partial_t B^i     =  F \partial_t \tilde{\Gamma}^i - \eta B^i 
$$
with $F$, again, given by equation~(\ref{eq:F_shift}). This has been
used in previous studies of head-on
collisions~\cite{Alcubierre02a,Alcubierre2003:BBH0-excision}. We
expect that this small difference in gauge will not lead to significant
differences in merger times.}.
The gauge parameter choices were $p=1$, $n=4$ and $\eta\in[2.8,3.5]$,
with the larger values of $\eta$ being used for the larger separations.
We do not expect the small difference in shift condition to affect the
results significantly.  In Fig.~\ref{fig:merger_times_others}, the
results of the QC simulations are compared with both Newtonian infall
times (dashed line) and relativistic head-on collisions (filled
diamonds).  The latter two results lie close together on essentially
parallel curves, agreeing with the results of~\cite{Anninos94b}, and
providing a baseline for coalescence times of purely plunging, head-on
collisions.

The lower curve in Fig.~\ref{fig:merger_times_others} indicates the
proper time to common AH formation, $\tau_\text{AH central}$, as
measured by an observer at the origin of the coordinate system. As
this is a symmetry point of the problem, this proper time is
invariantly defined and easily calculated by integrating the lapse,
and removes most of the potential coordinate ambiguity that might be
present in the other time determinations.  This proper time
$\tau_\text{AH central}$ is lower than the coordinate time because it
is measured in a region where the lapse is somewhat collapsed.

It is important to point out that the results obtained for the full
near-ISCO QC sequence are basically parallel to the curves for head-on
collisions.  Further, the proper time measured by observers at the
origin, shown by the squares, are also consistent with these results,
strengthening the argument that gauge effects are not responsible for
the trends.  Taken together, these results suggest that for the near
ISCO sequence studied here, the BH's are not in fact in quasi-circular
orbits, but are rather in an extended plunge towards coalescence.
Even at nearly twice the distance of the ISCO (QC-4), there is no
clear trend towards an orbit.  On the contrary, as shown in
Fig.~\ref{fig:merger_times_orbit}, the configurations merge in
progressively smaller fractions of an orbit.

\begin{figure}[t]
  \centerline{Part~(a)}
  \setlength{\unitlength}{1mm}  
  \begin{picture}(75,50)
  \put(-0.5,0){\includegraphics[scale=0.56]{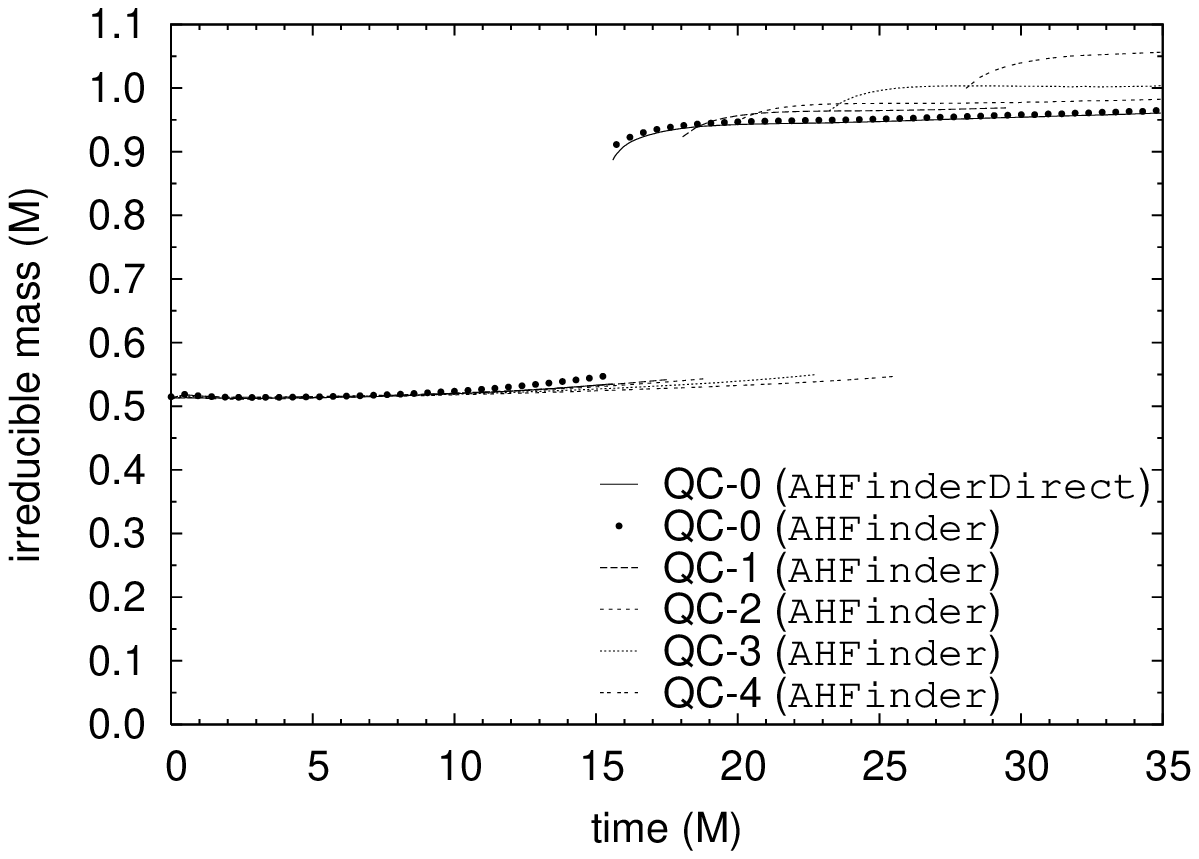}}
  \end{picture}
  \hbox{}\\[6mm]
  \centerline{Part~(b)}
  \begin{picture}(75,60)
  \put(-5,0){\includegraphics[scale=0.56]{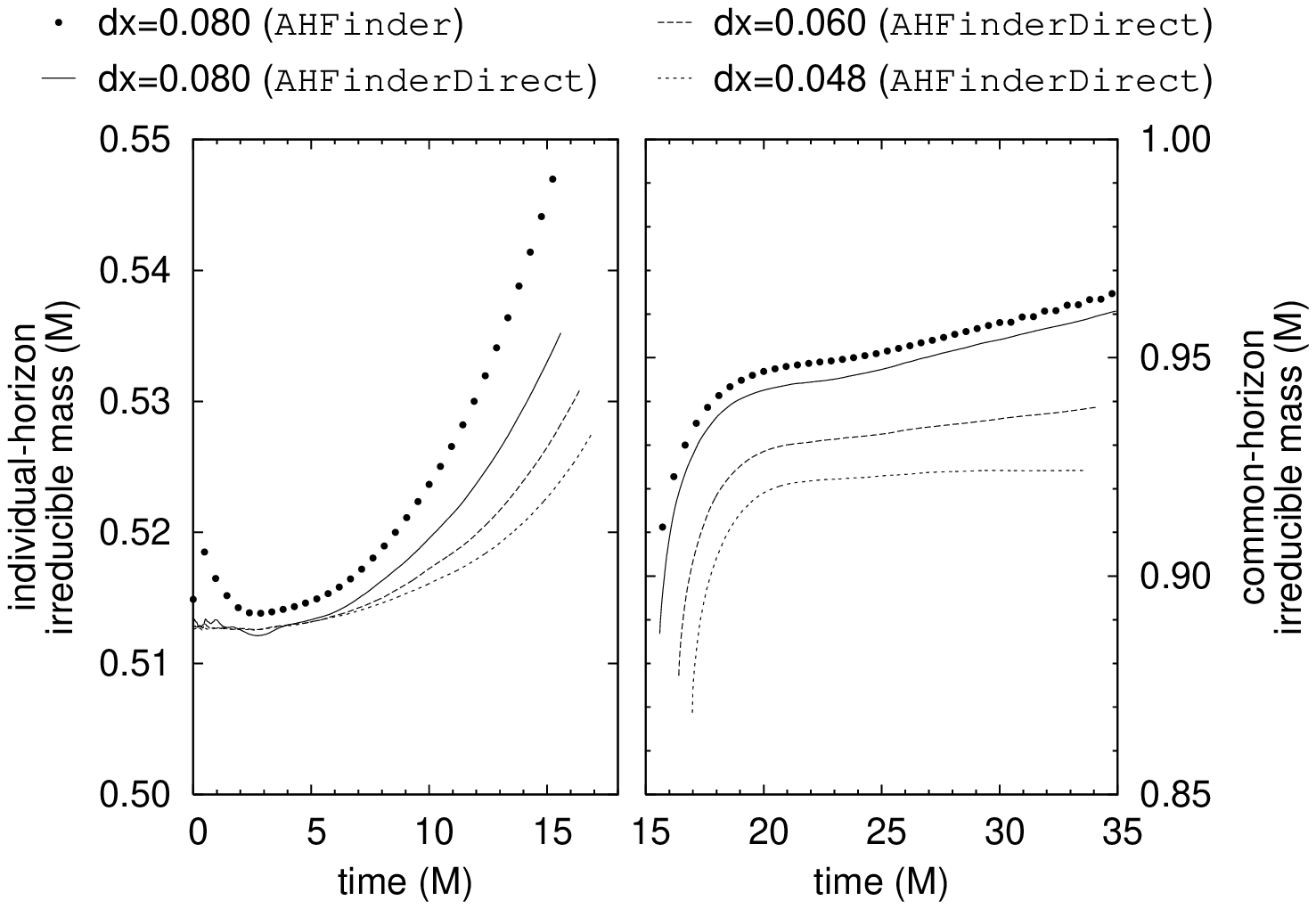}}
  \end{picture}
  \caption{Evolution of the irreducible mass $M_{\text{irr}}$ of the
    AH as a function of time.  Part~(a) shows the mass for each QC-N
    model, for the low resolution $dx=0.08$ evolutions, showing both
    individual- and common-horizon masses on the same scale.  (For the
    QC-4 model, the individual AHs are briefly ``lost'' due to the
    proximity of the excision region, somewhat before a common horizon
    is found.)  Part~(b) shows the mass for the QC-0 model for the
    $dx=0.08M$, $0.06M$, and~$0.048M$ evolutions, with
    individual-horizon masses in the left sub-plot and common-horizon
    masses in the right sub-plot.  For the $dx=0.08M$ QC-0 case,
    results from both apparent-horizon finders are shown,
    \texttt{AHFinderDirect} as a solid line, \texttt{AHFinder} as
    points.}
  \label{fig:ah_mass}
\end{figure}

\emph{Studies of Horizon Physics.}  Much can be learned about the
physics of the merger process, and of the final BH formed, through a
detailed study of the horizons (see,
e.g.~\cite{Anninos94f,Brandt94c,Masso95a,Alcubierre00b,Seidel99b}).
We begin by studying the masses of the AHs as a function of time for
each member of the sequence.  The extent to which horizon masses are
consistent with other quantities in the spacetime provides a
meaningful and physical measure of the overall accuracy of the
simulation.  As a measure of the BH mass, we study the irreducible
mass $M_{\text{irr}}=\sqrt{\text{Area}/16\pi}$.  In
Fig.~\ref{fig:ah_mass} we plot $M_{\text{irr}}$ for each model in the
near-ISCO sequence over the course of evolution.  Initial individual
horizon masses of approximately $0.5 M_\text{ADM}$ are found to remain
essentially constant, until the formation of a common horizon.  After
merger, we find that in all cases the results are consistent with the
ADM mass, and again hold a rather constant value in the subsequent
evolution.  Errors in the calculated mass are greater for the late
merging cases, but even in the worst cases the masses do not exceed
the ADM mass by more than about 5\%.  Drift in the final horizon mass
is quite small, even at late times when errors accumulate and outer
boundaries might be expected to have an influence.  Results from the
two different apparent-horizon finders are very similar, providing a
further consistency check.

Figure.~\hbox{\ref{fig:ah_mass}(b)} shows that these results are only
weakly resolution-dependent.  At higher resolutions, we see slightly
smaller horizon masses, somewhat later merger times, and a much
smaller upward drift of the common-horizon mass.  Unfortunately, these
results do not show quantitative convergence. This appears to be due
to insufficient resolution, particularly in our coarsest ($dx =
0.08M$) models.  (Comparing this result to the good second~order
convergence of the Hamiltonian constraint at comparable resolutions
(Fig.~\ref{fig:ham_convergence}), it would appear that the
apparent-horizon diagnostics require higher resolution than the
Hamiltonian constraint to reach the quantitative-convergence regime.)

Fig.~\hbox{\ref{fig:ah_mass}(b)} shows that the individual-horizon
masses are slightly greater than ${\textstyle \frac{1}{2}} M_\text{ADM}$.
This is not surprising, because we expect a (negative) gravitational
binding energy of the two black holes in the initial data, although
there is also orbital kinetic energy and possible gravitational
radiation in the initial slice not taken into account.

We note that we have not used the traditional measure of mass for a
black hole,
\begin{equation}
M_\text{AH} = \sqrt{M_\text{irr}^2 + \frac{J^2}{4M_\text{irr}^2}}
\label{AHmass}
\end{equation}
which includes the angular momentum (and adds a small contribution to
the mass), because the angular momentum of the final BH is not known
in advance~\cite{Smarr73a,Christodoulou70}.  Some previous
studies~\cite{Brandt94c, Alcubierre00b,Seidel99b} could safely assume
that very little or no angular momentum was radiated, and were
therefore able to use angular momentum of the spacetime to compute
this mass, and then to calculate the total energy radiated.  For the
cases studied here, the angular momentum of the final black hole must
be measured in each case, as we discuss below.

Beyond simple determination of the AH masses, much more physics of the
merger process can be obtained by studying the dynamics of the merged
horizons.  In this study, in all cases the black holes could be
evolved to well past common apparent horizon formation, allowing
numerous physical characteristics of the final black hole to be
computed as it settles.

\begin{figure}
  \hspace{2pc}\includegraphics[width=18pc,height=12pc]{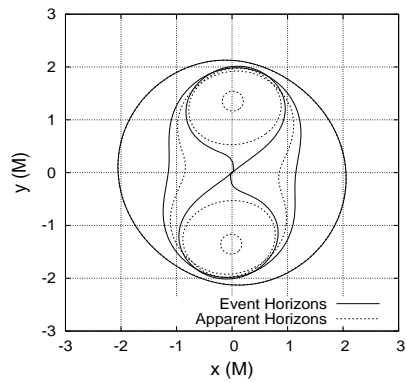}
  \caption{Event and apparent horizons for an evolution of the QC-1
  model ($dx=0.08 M$), showing horizons on the initial slice ($t=0.0
  M$), when the EH first merges ($t=13.0 M$), at the appearance of the
  first common AH ($t=18.1 M$), and at a late time slice ($t=30.0 M$)
  when the system has essentially settled and the two contours cannot
  be distinguished anymore.}
   \label{fig:eh_ah}
\end{figure}

We begin with a study of the shape of the horizons.  First, we note
that for evolutions that can be carried out to a point of near
stationarity, when the final black hole has settled down, the event
horizon can be accurately traced by integrating null surfaces backward
in time through the spacetime~\cite{Anninos94f,Diener03a}, and further
that a detailed analysis of the physics of these event horizons can be
carried out~\cite{Masso95a,Matzner95a,Caveny-Anderson-Matzner-2003a}.
We have carried out the
analysis for these 3D spacetimes using the method described
in~\cite{Diener03a}.  In Fig.~\ref{fig:eh_ah}, we show a comparison of
the coordinate locations of the event and apparent horizons for the
near-ISCO case QC-1 at various times for a resolution of $dx=0.08M$.
On the initial slice $t=0$, no common event or apparent horizon
exists.  Due to the small size of the horizons (compared to the grid
spacing), the location of the event horizons can not be found with
sufficient accuracy on the initial slice. For that reason we only show
the location of the initial apparent horizons. By time $t=13M$, the
event horizon sections have just merged, with an interesting caustic
structure that will be studied elsewhere.  Interestingly, away from
this caustic region the coordinate locations of the EH and AH agree
with great precision, with the EH just outside the AH.  By time
$t=18M$, we find the first appearance of a common AH, which as
expected occurs later than the merger of the EH.  Finally, at the late
time $t=30M$, the AH and EH locations agree very well, indicating the
final merged black hole has settled down.  The close agreement in
location of both local temporal measures (AH) and global temporal
measures (EH) of horizon structures gives confidence in the overall
accuracy of the merger evolution.

The coordinate location of horizon structures are useful for
comparisons and for topology measures, but they say nothing about pure
geometric measurements from which physical quantities can be derived.
As shown first by Smarr~\cite{Smarr73b} for a Kerr BH, the spin
parameter $a/m$ uniquely determines the geometric shape of the BH
horizon.  As discussed in~\cite{Anninos94f,Brandt94c}, this relation
allows one to use the horizon shape as a tool to measure the spin of a
black hole in dynamic
spacetimes~\cite{Anninos94f,Brandt94c,Seidel99b}.  A reasonable
measure of the shape of BHs that are nearly Kerr is the ratio of
proper horizon circumferences measured along polar, $C_p$, and
equatorial, $C_e$, geodesics~\cite{Brandt94c}.  The value of
$C_r=C_p/C_e$ is unity for a Schwarzschild BH, and less than unity for
Kerr BHs, with the precise value depending uniquely on the rotation
parameter $a/m$ as discussed below.  As shown in~\cite{Brandt94c}, a
distorted rotating BH should oscillate about its equilibrium Kerr
shape.  In axisymmetry, the polar circumference is uniquely defined,
but in full 3D one has multiple choices.  In this study, we compute
the ratio $C_r$ along circumferii in the coordinate planes (because of
symmetry about the plane of the orbit, the equatorial circumference is
uniquely defined).

\begin{figure}
  \vspace{3mm}
  \includegraphics[width=20pc]{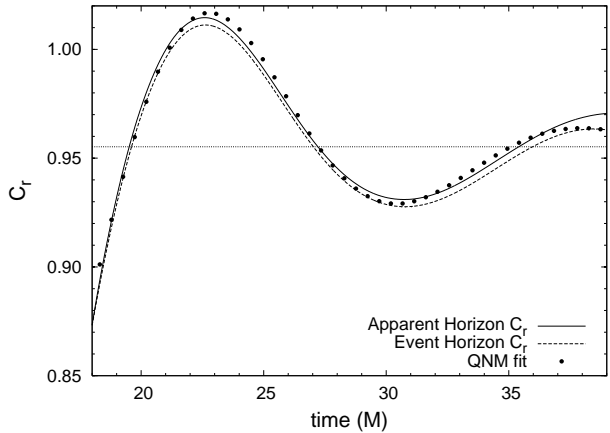}
  \caption{Comparison of EH to AH geometry for case QC-1.  The ratio
  of the polar and equatorial horizon circumferences, $C_r=C_p/C_e$,
  after formation of common horizons, measured in the \emph{xz} plane,
  is shown for the low resolution ($dx=0.08 M$) case. Both AH and EH
  horizons oscillate at the QNM frequency of the final Kerr BH, as
  confirmed by the fit shown (see text).}
  \label{fig:horizon_oscillation}
\end{figure}

Fig.~\ref{fig:horizon_oscillation} shows results for the case just
outside the ISCO, QC-1, plotting the evolution of the ratio of polar
to equatorial circumference $C_r$, measured in the \emph{xz} plane,
for both common apparent and event horizons, for the low resolution
case with $dx=0.08M$.  The EH and AH shapes agree extremely well, as
was also suggested by Fig.~\ref{fig:eh_ah}.  Both curves reveal the
oscillation pattern expected in the geometry of a distorted rotating
BH formed in the merger.

Following~\cite{Brandt94c,Seidel99b}, we use the oscillation pattern
in $C_r$, and its equilibrium value, to extract important information.
The final BH formed during a merger is expected to have its
quasi-normal modes (QNMs) excited.  For each member of the Kerr
family, characterized by $a/m$, a set of discrete QNMs exists.
Further, for each $(\ell,m)$ pair, there is a fundamental mode and
higher overtones.  In principle, all of these mode may be excited, but
in practice the higher overtones are much more strongly damped than
the fundamental.  Although associated with the potential barrier
surrounding the BH, the QNMs excite gravitational waves that propagate
both to infinity and also inward, crossing the horizon where they
manifest as oscillations in the horizon
geometry~\cite{Anninos94f,Brandt94c}.  Therefore, one should find
various QNMs in the oscillations of $C_r$, and can use this
information to determine the angular momentum of the final BH.

The QNMs for rotating black holes have been studied extensively via
perturbation theory, and have been computed as a function of $a/m$ in
Refs.~\cite{Leaver86,Seidel90a}.  Using these known values, and
building on previous work~\cite{Brandt94c,Seidel99b}, we have devised
the following procedure to analyze the horizon oscillations.  One can
show analytically that for a Kerr black hole, the quantity $C_r$ has a
value given by
\begin{equation}
C_r = \frac{1+\sqrt{1-(a/m)^2}}{\pi} \, 
        E \left( -\frac{(a/m)^2}{(1+\sqrt{1-(a/m)^2})^2} \right) ,
\label{eq:Cr-exact}
\end{equation}
where $E(k)$ is the complete elliptic integral of the second kind
\[
E(k)=\int_0^{\pi/2}\sqrt{1-k\sin^2\theta}\,d\theta.
\]
For simplicity $C_r$ can be well approximated by the
empirical-theoretical formula, accurate to within less than
1\%~\cite{Brandt94c},
\begin{equation}
C_r = \frac{\sqrt{1-(a/m)^2}+1.55}{2.55}.
\label{Cr}
\end{equation}
A perturbed BH horizon with a given value of $a/m$ oscillates about
this equilibrium value, largely through a superposition of its
QNMs~\cite{Anninos94f,Brandt94c}.  Therefore, the function $C_r$
should be composed of an equilibrium value and a linear combination of
various QNMs~\footnote{As shown in~\cite{Anninos94f,Brandt94c}, the
horizons oscillate at the tabulated QNM frequencies without need to
account for differences in proper time caused by commonly used
singularity avoiding slicings, as expected.  This can be understood as
a canceling of the blue-shift of the waves crossing the horizon and the
slowing of time by the slicing condition.}.  We use an iterative
least-squares fitting procedure to determine the value of $a/m$ that
best fits the extracted curve $C_r$.
   
In these studies, we have found it is sufficient to consider only the
offset, and the phase and amplitude of the fundamental modes (given by
$\ell-m$) as fitting parameters.  Here we have used the
$\ell=2,m=0,1,2$ modes, but we note that unless $a/m$ becomes very
large, the QNM frequencies are all very similar to each other.  We
begin with a fiducial set of modes for a given $a/m$ value, adjust
their phase and amplitude and the offset value, which determines the
approximate value of $a/m$ through Eq.~(\ref{Cr}) for the first
iteration.  We then use this value of $a/m$ to choose the QNM
frequencies for the next iteration, which through the offset
determines the new value of $a/m$.  For the data sets studied here,
this procedure converges extremely rapidly, independent of the
starting value of $a/m$. We have also used a binary search technique
over $a/m$ to find a minimum error in the least squares
procedure. This method gave identical final results for $a/m$. We note
that the fitting procedures use the QNM frequencies for unit BH mass
and vary only $a/m$. This is justified by the fact that
$M_\text{BH}\approx M_\text{ADM}\approx 1$.

Turning back to Fig.~\ref{fig:horizon_oscillation} for QC-1, we see
that not only do the curves obtained for both the EH and AH closely
agree, they both show the oscillations of the final Kerr BH formed in
the coalescence process.  The result of the fitting procedure to the
Kerr modes, described above, is also displayed.  We see a beautiful
agreement between both the event and apparent horizon measurements,
which are completely independent, and the fit to the QNMs of the final
Kerr BH.

This result shows conclusively that in the coalescence process, the
QNMs of the final black hole are strongly excited, and dominate the
evolution of the horizon after its formation.  Although we have not
attempted to extract waves in the far zone in these studies, due to
the nearness of the outer boundary, this result also implies that at
late times QNMs will dominate the waveforms.  Below, we study the
determination of the spin of the final black hole from this procedure.

In many cases we are able to evolve these systems significantly beyond
the formation of the final BH, so that its properties and be studied
in depth.  We then have additional analysis tools at our disposal. For
models where an event horizon was located (QC-0 and QC-1), the horizon
generators can in principle provide an independent estimate of the
spin of the final black hole, as shown
in~\cite{Anninos94f,Libson94a,Masso95a}. Several factors complicate
the analysis of the horizon generators in these cases. First of all,
for QC-0 and QC-1, the evolutions did last long enough past merger for
us to track the EHs accurately during the merger, but not quite long
enough to track them accurately at the time when the black hole has
almost settled down to its final stationary configuration. It is only
at this late time that the formula relating the generator angular
velocity to $a/m$ is correct. Additionally, the angular velocities are
gauge dependent. Therefore we have to correct the coordinate angular
velocities not only for the effects of our co-rotating coordinate
system but also for any co-rotation shift developed by the
$\Gamma$-driver shift (equations~(\ref{eq:shift})
and~(\ref{eq:F_shift})).  We have, as yet, not found a satisfactory
way of doing this and are consistently getting lower estimates for
$a/m$ using the generators compared to the results obtained by fitting
QNMs to the EH circumference ratios.

The isolated horizon formalism~\cite{Ashtekar99a,
Ashtekar-etal-2002-dynamical-horizons} provides an additional tool to
study the final horizon to extract physical information.  In this case
we apply this formalism to the final merged AH to determine a spin
by finding an approximate solution to the Killing equation in the
surface~\cite{Dreyer-etal-2002-isolated-horizons}. While such a
solution might not exist immediately after the formation of the common
AH we have found that within $7-10M$ the formalism provides reliable
information as shown by the $a/m$ measure approaching a constant value.

\begin{figure}
  \vspace{3mm}
  \includegraphics[width=20pc]{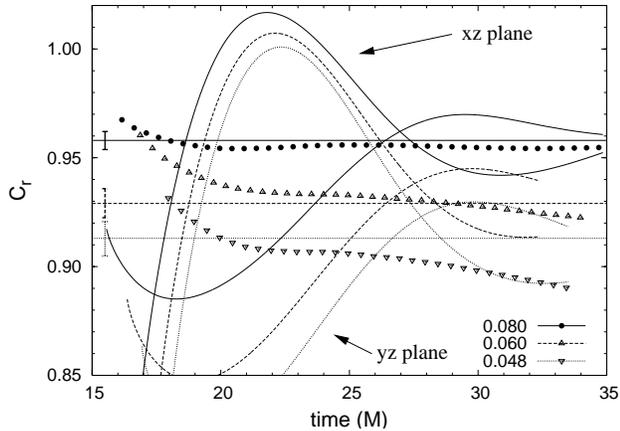}
  \caption{Comparison of $a/m$ measurements for case QC-0. We show the
  circumference ratio $C_r$, measured on the AH, for 3 different
  resolutions for the two different polar circumferences (computed in
  the $xz$ and $yz$ plane). We also show the results of the fitting
  procedure to determine an ``offset'' value of $C_r$ (straight lines)
  for the average of the measurements of circumference ratios as
  listed in Table~\ref{tbl:aMcmp}, showing good agreement with the
  curves. The offsets determined from the two individual fits are used
  to compute the error bars at the left side of the figure.  Also
  plotted is the result from the isolated horizon computation
  converted into a circumference measure for comparison. We note that
  the value from the isolated horizon has developed a plateau between
  $t=20-25M$, which agrees relatively well with the fits to $C_r$ for
  the geometric measurements.}
  \label{fig:spin_resolution}
\end{figure}

We now apply both the horizon geometry and isolated horizon
measurements discussed above to the case studied in most depth for
this paper, QC-0.  First, as noted above, the geometric measurements
$C_r$ have been made in two different ways, using the polar
circumference in both the $xz$ and $yz$ planes.  In
Fig.~\ref{fig:spin_resolution} we show the evolution of $C_r$ in both
of these planes, as well as a fit to determine the ``offset'' value,
which allows us to estimate the spin of the final BH as described
above.  For comparison, we have taken the measurement of the isolated
horizon spin, and converted it into an instantaneous $C_r$ via
Eq.~(\ref{eq:Cr-exact}).  We see that this measurement settles into a
plateau after about $t=20M$, agreeing fairly well with the geometric
measurements.

In Table~\ref{tbl:aMcmp} we turn these results into quantitative
measurements of the spin of the final BH, for different methods and
different resolutions, all for the case QC-0.  First, we note that at
the lowest resolution, all measurements give consistently low values
of the final spin, averaging to $a/m_\text{lowres} \approx 0.46$.
This estimate seems extremely low, however, for a system with initial
$J/M^2=0.78$.  However, the study of resolution effects on this
quantity shows that it is very sensitive to resolution.  At the
highest resolution studied, we measure the final BH to have spin
$a/m_\text{hires} \approx 0.65$.  Unfortunately, the various measures
used here do not show clear second order convergence properties, and
therefore it is not possible to perform a Richardson extrapolation.

Therefore, with present resolution and techniques, we can say with
confidence that low resolution simulations underestimate the spin of
the final black hole, and that at the highest resolutions possible in
the present study, we believe we are still underestimating the final
spin by a small amount.  A separate investigation of 3D evolution of
single, spinning, distorted black holes is underway to investigate
resolution and computational parameters needed to accurately resolve
and determine the spin from horizon
measurements~\cite{Diener2004:dist-bh-rotating}.  These results do
indicate that the merger process, for the cases studied, radiates a
small amount of angular momentum (less than about 20\%).  The results are
summarized in Table~\ref{tbl:JErad}.

\begin{table}
  \begin{ruledtabular}
  \begin{tabular}{cccccc}
     & \multicolumn{4}{c}{\dotfill{~$a/m$ determined from~}\dotfill}  \\
  $dx$ & $C_r$ $xz$  & $C_r$ $yz$ & $C_r$ avg  & IH                   \\
  \hline 
  0.080 & 0.429 & 0.471 & 0.450 & 0.470 \\
  0.060 & 0.548 & 0.597 & 0.573 & 0.570 \\
  0.048 & 0.603 & 0.653 & 0.628 & 0.660 \\
  \end{tabular}
  \end{ruledtabular} 
  \caption{Table comparing different angular momentum measures $a/m$
  for the final black hole in a QC-0 merger simulation at different
  resolutions $dx$. $C_r$ refers to $a/m$ determined via a QNM
  fit as described in the text. The cases \emph{xz}, \emph{yz}
  refer to the ratio of the circumference in the \emph{xz}
  to the \emph{xy} circumference and of the \emph{yz} to the \emph{xy}
  circumferences. The \emph{avg} denotes the average of the
  two.  \emph{IH} denotes the value computed by the isolated
  horizon finder, measured on the plateau at $t=25M$.}
  \label{tbl:aMcmp}
\end{table}

\begin{table}
  \begin{ruledtabular}
  \begin{tabular}{cccccc}
  $dx$ & 0.080 & 0.060 & 0.048 \\
  \hline 
  $a/m$                & $0.450\pm0.021$  & $0.572\pm0.025$ & $0.632\pm0.028$ \\
  $M_\text{irr}$       & 0.947            & 0.933           & 0.923           \\
  $M_\text{AH}$        & $0.973\pm0.003$  & $0.978\pm0.005$ & $0.980\pm0.006$ \\
  $J_\text{rad}$ (\%)  & $45.3\pm2.9\phantom{0}$
                                          & $29.6\pm3.7\phantom{0}$
                                                    & $22.1\pm4.5\phantom{0}$ \\
  $E_\text{rad}$ (\%)  & $3.61\pm0.25$    & $3.12\pm0.45$   & $2.97\pm0.59$   \\
  \end{tabular}
  \end{ruledtabular} 
  \caption{The table shows the radiated angular momentum and energy
  for the different resolutions $dx$ for the case QC-0. Starting from
  Table~\ref{tbl:aMcmp} we use the minimum and maximum values for
  $a/m$ at the different resolutions.  The ranges given refer to these
  values for $a/m$. For the irreducible mass $M_\text{irr}$ we take
  the values reported from the apparent horizon finder at $t=25M$. We
  compute the horizon mass $M_\text{AH}$ which includes angular
  momentum and then compute radiated energy and angular momentum using
  initial data values of $M_\text{ADM}=1.01$ and
  $(J/M^2)_\text{ADM}=0.779$.}
  \label{tbl:JErad}
\end{table}

Finally, we turn to the computation of the final black hole mass. In
Table~\ref{tbl:JErad} we also summarize computations of both the
irreducible mass (not including spin) $M_\text{irr}$ and the horizon
mass (including spin) $M_\text{AH}$ as defined in
Eq.~(\ref{AHmass})~\footnote{We use $J=a/m M_\text{AH}^2$, and solve
the resulting equation for $M_\text{AH}$.}.  In computing
$M_\text{AH}$ we use a range of values for $a/m$, as listed in
Table~\ref{tbl:aMcmp}, to compute the error bars listed in
Table~\ref{tbl:JErad}.  Note that there are additional errors in
particular regarding the value of $M_\text{irr}$ which have not been
included.  We compute
$E_\text{rad} = (M_\text{ADM} - M_\text{AH})/M_\text{ADM}$ and
$J_\text{rad} = (J_\text{ADM} - (a/m) M_\text{AH}^2)/J_\text{ADM}$.
Here we see that, although we are unable to
precisely determine the total energy radiated with present techniques,
we nonetheless have demonstrated the high degree of accuracy that is
now achievable for BH merger simulations.  Our highest resolution
simulations find a final BH mass within 3\% of the ADM mass, and the
trend as a function of resolution indicates the total energy radiated
to be probably less than this.

We note that Ref.~\cite{Baker:2002qf} estimated, via completely
different methods, that roughly 12\% of the angular momentum is
radiated, for the same initial data sets studied here, and that the
total energy radiated is 3\%.  These results are consistent
with those that can now be obtained via full nonlinear evolution.
These points are discussed further below.


\section{Discussion}
\label{sec:discussion}

We have performed the first systematic study of binary black holes
constructed to be in approximately quasi-circular orbits via effective
potential methods, using the full machinery of 3D numerical
relativity.  The data sets chosen range from the innermost circular
orbit, or ISCO (QC-0 in our notation), as determined from this
construction, to data sets with separations significantly beyond this,
in what we refer to here as the near-ISCO regime.  In each case, using
a combination of recently developed techniques including excision, new
gauges, the BSSN formulation, and a co-rotating frame, we have been
able to carry out the evolutions until after a common apparent horizon
formed, indicating an upper bound on the time of merger of the two
black holes.

The coordinate times at which the common horizon formed were found to
be significantly less than the timescale of an orbit, defining this
timescale from the initial angular velocities. We also show that as
one evolves initial data sets further out in this near ISCO regime,
the holes merge after longer times but for progressively smaller
fractions of an orbit. We compared these results with results from
head-on collisions from corresponding distances, and found that the
time to merger shows a very similar trend. This suggests that the
initial data models studied here are already in an extended plunge,
and not in quasi-circular orbits as one might expect from the
effective potential considerations.  We also found that the angular
velocity of the co-rotating frame needed to keep the holes in a
``head-on'' collision as seen in coordinate space, is roughly 2/3 of
what would be expected from considerations of the initial data, even
after one takes into account the intrinsic co-rotating component of
the shift condition used~\cite{Alcubierre2003:co-rotating-shift}.
This result is consistent with our findings that the holes are merging
much more rapidly than expected from initial data considerations.

It is important to point out that these results are also consistent with
two completely different studies, the Lazarus results about the near
ISCO regime, and investigations of binary neutron star systems near
the ISCO.

The same black hole data sets that we used here have been studied in the
hybrid perturbative-numerical Lazarus
approach~\cite{Baker00b,Baker:2001nu,Baker:2001sf}. This approach
evolves the same data sets numerically, only to a ``linearization''
time, of roughly $10M$, where the two holes are close enough that the
system could be treated as a single perturbed BH, after which a
perturbative approach was used to complete the evolution.  Recall the
results from Lazarus on the linearization time along the initial data
sequence.  According to Ref.~\cite{Baker:2002qf}, for the cases
studied here, QC-0 to QC-4, the linearization time increases with
separation, but with a slope similar to, or even less than, that seen
for coalescence times found here.  This result gives support to the
notion that this black hole sequence is not approaching an orbital
configuration out to QC-4.

Simulations of inspiral sequences of binary neutron stars were
carefully studied by Miller and Miller et al.\
in~\cite{Miller03c,Miller:2003vc}, who examined the correlation of
waveforms from model evolutions against PPN waveform templates.  These
authors find strong evidence that the circular orbit assumption can
give rather poor approximations for two neutron stars in the near ISCO
regime. Extrapolating these results to the binary BH case, note that
the initial separations of QC-0 to QC-4 clearly fall into the regime
where, based on~\cite{Miller03c,Miller:2003vc}, one would not expect
an orbital configuration. Again, this is consistent with our results
that indicate a plunging configuration for all data sets studied
here.

A number of issues surrounding the merger time numbers that we have
presented need to be considered with care, in particular effects due
to boundary treatment, numerical resolution, and the gauge choice.

We have taken some care to assure that boundary effects do not
significantly alter the orbital dynamics and merger times
picture. Boundaries have been placed at a distance causally separated
from the common horizon, although even without this restriction we
have obtained essentially identical runs with boundaries in both
closer and further separations, indicating that they do not strongly
influence the local dynamics of the black hole horizons for the
timescales considered here. Though the numerical domain of dependence
is larger, the Hamiltonian constraint was found to converge to second
order in the region of the common horizon.  We also found that the
merger times are not significantly affected by the resolutions used in
these studies.

As a point of reference, in Ref.~\cite{Miller:2003vc},
high resolution (0.054\,$M$) was required to get convergent
results. Our coarsest resolution of 0.08\,$M$ is somewhat lower than
this, but our medium and high resolutions (0.06\,$M$ and 0.048\,$M$)
are comparable.  Unlike~\cite{Miller:2003vc}, we were able to ensure
that the outer boundary remains causally disconnected from the merger
region for all models studied. As a result, we found that changes to
the outer boundary location had only small effects on the time of
merger (well within the error bars of
Fig.~\ref{fig:merger_times_orbit}).

As remarked above, our studies indicate that for these BH
simulations resolution and boundary location effects are not
responsible for the quick coalescence times, especially for the cases
closer to the ISCO (e.g., QC-0,1).

In addition to resolution and boundary effects, gauge effects will
have an influence on the time at which a common AH can first be found.
In order to remove some potential slicing and coordinate ambiguity,
for all cases we computed the proper time to common AH formation along
the origin, which is at all times a point of symmetry between the
holes.  This measure of time is actually gauge invariant for EHs, but
differs from time measured at infinity.  The results show the same
trend as above, that the holes are in an extended plunge and the
greater separations do not show any tendency towards approaching an
orbit.

As described above, we have used a consistent set of gauge parameters
for all the reported results (except the EH results). Variations of
these parameters (to the extent that was possible in order to still
achieve a long-lasting run) were found to vary the time to appearance
of a common AH by only a small amount, within the displayed error
bars. However, in the initial $5M$ of evolution when the black holes
are still quite small on the grid, resolution effects can influence
the rate at which the gauges evolve (collapse of the lapse, for
instance), resulting in an offset even with identically chosen gauge
parameters. The error bars displayed in
Fig.~\ref{fig:merger_times_orbit} are not statistically derived, but
rather indicate the greatest range of results from many experiments
with resolution and gauge parameters.  We have not, however, observed
a significant increase in the time to common apparent horizon
formation for any of the studied models. In particular for the QC-0
and QC-1 cases where a large number of runs under various conditions
of gauge and resolution were carried out, we do not expect the merger
time to vary greatly with changes to computational parameters (gauge,
resolution, boundary).  

On the other hand, as indicated by the growing error bars, 
there is increasing uncertainty in the QC-3 and QC-4 cases.  
We note, for example, that the fisheye coordinate
transformation that stretches the coordinates in physical space is very close
to the initial puncture locations in these cases.  This, or additional gauge
effects, could conceivably
affect orbital evolution properties.  Further experiments for initial data
in this regime and beyond are underway with
mesh refinement, and with a wider range of gauges, 
and will be reported in future publications.

We also applied a number of tools to measure various physical
quantities from the final BH horizons formed in the merger, which were
found to agree well. Both event and apparent horizons were determined
for the QC-0 and QC-1 cases~\footnote{Event horizons for the other
cases would have required longer runtimes, or deeper excision regions,
necessitating a parameter study of available gauges.}.  Geometric
measures of the EH and AH horizon geometry were used to compute the
mass and spin of the final BH, as well as the oscillation frequency of
the radiation.  In all cases, the final BH mass was found to be close
(within 4\%) to the ADM mass of the spacetime, providing a global
measure of the accuracy of the simulations.  Appearing with a delay of
$4-5M$ from the first joining of the EH, the common AH was found to
coincide with the EH extremely well very soon after it merged.  As we
discuss below, in the highest resolution studies of the case QC-0,
going beyond overall consistency of energy measures, we are nearly
able to predict the total energy radiated.

We also used the isolated horizon formalism to study the dynamics of
the final BH, especially the final spin, and obtained similar results.
It is interesting that within a few $M$ of common AH formation, the
estimate of the final angular momentum of the merged black hole from
the AH distortion (via $C_r$, defined above) and the same measure of
EH distortion, agree well with the independent estimate obtained by
the isolated horizon formalism.  This suggests that it takes only a
very short time for the final black hole to settle to a sufficiently
Kerr-like state so that each of the approaches can be used.  The fact
that three very different measures of the spin of the final black hole
agree so closely with each other gives us confidence in the validity
of the results.

However, the low resolution evolutions consistently under-predict the
estimated spin by a significant amount for the cases studied.  The
resolution studies indicate that at higher resolution, the spin and
mass measurements of the final BH horizon approach the asymptotic ADM
values, indicating that total radiation is small.  For the QC-0 case
studied in depth, we estimate that roughly 20\% or less of the total
angular momentum is lost in the merger process.  It is important to
note that although the spin of the final black hole is sensitive to
the resolution, the coalescence time is not.  Even for the more
distant configurations in the pre-ISCO family studied here, resolution
affects the coalescence time only relatively weakly, and the highest
resolution studies find the holes to coalesce well before an orbital
time period even for QC-4.

We singled out the ``ISCO'' case, QC-0, for very detailed study.
In addition to the calculations that indicate the system to radiate
of order 20\% or less of its angular momentum, we also
studied the radiated energy as a function of resolution.  By taking
into account the estimated angular momentum of the final black hole
formed in the merger, we were able to compute the total mass of the
final BH, $M_\text{AH}$, defined by Eq.~(\ref{AHmass}).  In all
cases, it was found to be within 4\% of the ADM mass, and at the
highest resolution we found results consistent with 3\% or less of
total energy being radiated.

Note that the results of the Lazarus
approach~\cite{Baker:2001nu,Baker:2002qf} for the total energy and
angular momentum radiated are consistent with our fully numerical
results for the same initial data.  The fact that the results are
similar, yet the evolutions were carried out with completely different
evolution equations (ADM vs. BSSN) and gauge conditions (maximal/zero
shift vs. 1+log/Gamma-driver shift), and then were further
differentiated by the linear and nonlinear approaches, gives added
confidence in the results.

The primary focus of this paper has been on the orbital dynamics of a
family of binary black holes expected to be in quasi-circular orbits,
and not on the waveforms generated.  However, by making detailed
measurements of the apparent and event horizon geometries, we have
been able to show conclusively that the quasi-normal modes of the
final black hole are strongly and cleanly excited, and they are
dominated by the fundamental $l=2$ modes.  These QNMs will therefore
be very strongly seen in the waveforms far from the holes.

While the simulations reported in this paper were being analyzed for
publication, another study was carried out using similar techniques,
but with a different code~\cite{Bruegmann:2003aw}.  In that study,
another binary black hole data set, from a family which is similar to
the family studied here but with $L/M =9$, further beyond the ISCO
than the present study (QC-4 has $L/M=7.84$), was found to evolve
beyond the predicted orbital time period ($114M$) without finding a
common apparent horizon.  Although our results show a trend for this
family of data sets to be in near plunge when $L/M < 8$, the results
in~\cite{Bruegmann:2003aw} for $L/M=9$ are not necessarily in
contradiction with the findings of this study.  Further, there are a
number of differences in the methodology, including gauge conditions
and the computational grid setup.  In this paper, we have stressed the
importance of sufficient resolution to determine accurately the
angular momentum of the final black hole.  At the same time, we have
not seen a strong dependence of the coalescence time on the resolution
for the range of simulations carried out here.  We note that although
our highest resolution runs are for $dx=0.04M$, slightly coarser than
the typical fine grid resolution $dx=0.03125M$ reported
in~\cite{Bruegmann:2003aw}, we would not expect this alone to account
for the differences.  Detailed study, beyond the scope of the present
investigation, will be required for data sets further into the
near-ISCO regime to investigate these results.

The question arises at which point of the QC sequence there is a
transition from plunge to inspiral. For sufficiently large separations
this type of initial data is known to be a good approximation to
quasi-circular post-Newtonian data~\cite{Cook94,Tichy:2003qi}. The
results presented here indicate that the effective potential method
does not produce data sets that will evolve in quasi-circular orbits
at initial proper separations less than $8M$.  We note that for a
different initial data set the ISCO was found at an approximate
distance of $L/M = 6.61$, much further out than with the method of
minimum effective binding energy used for this
study~\cite{Gourgoulhon02,Grandclement02,Damour:2002qh}.

Recent progress in the use of mesh refinement in our
codes~\cite{Schnetter-etal-03b,Bruegmann:2003aw}, as well as the
introduction of higher order finite difference operators, will enable
us to minimize resolution dependent effects. Work is in progress to
investigate the transition in the initial data between orbiting and
plunging black holes, and to study the sequence of initial data sets
developed using the HKV conformal thin sandwich
approach~\cite{Gourgoulhon02,Grandclement02,Damour:2002qh,Cook:2004kt}.


\acknowledgments

The authors would like to thank Erik Schnetter and Ryoji Takahashi for
many useful conversations and contributions. Results for this paper were
obtained using computing time allocations at the AEI, CCT, LRZ, NCSA,
NERSC, PSC and RZG. We use Cactus and the \texttt{CactusEinstein}
infrastructure with a number of locally developed thorns. This work was
supported in part by DFG grant ``SFB Transregio~7:
Gravitationswellenastronomie'', by NSF grants PHY-02-18750, PHY-02-44788
and PHY-03-54842, by the EU Programme `Improving the Human Research
Potential and the Socio-Economic Knowledge Base' (Research Training
Network Contract HPRN-CT-2000-00137), by NASA grant NNG04GL37G and by
DGAPA-UNAM grants IN112401 and IN122002.


\bibliographystyle{apsrev}

\bibliography{references}

\begin{thebibliography}{59}
\expandafter\ifx\csname natexlab\endcsname\relax\def\natexlab#1{#1}\fi
\expandafter\ifx\csname bibnamefont\endcsname\relax
  \def\bibnamefont#1{#1}\fi
\expandafter\ifx\csname bibfnamefont\endcsname\relax
  \def\bibfnamefont#1{#1}\fi
\expandafter\ifx\csname citenamefont\endcsname\relax
  \def\citenamefont#1{#1}\fi
\expandafter\ifx\csname url\endcsname\relax
  \def\url#1{\texttt{#1}}\fi
\expandafter\ifx\csname urlprefix\endcsname\relax\def\urlprefix{URL }\fi
\providecommand{\bibinfo}[2]{#2}
\providecommand{\eprint}[2][]{\url{#2}}

\bibitem[{\citenamefont{Alcubierre et~al.}(2003)\citenamefont{Alcubierre,
  Br\"ugmann, Diener, Koppitz, Pollney, Seidel, and Takahashi}}]{Alcubierre02a}
\bibinfo{author}{\bibfnamefont{M.}~\bibnamefont{Alcubierre}},
  \bibinfo{author}{\bibfnamefont{B.}~\bibnamefont{Br\"ugmann}},
  \bibinfo{author}{\bibfnamefont{P.}~\bibnamefont{Diener}},
  \bibinfo{author}{\bibfnamefont{M.}~\bibnamefont{Koppitz}},
  \bibinfo{author}{\bibfnamefont{D.}~\bibnamefont{Pollney}},
  \bibinfo{author}{\bibfnamefont{E.}~\bibnamefont{Seidel}}, \bibnamefont{and}
  \bibinfo{author}{\bibfnamefont{R.}~\bibnamefont{Takahashi}},
  \bibinfo{journal}{Phys. Rev. D} \textbf{\bibinfo{volume}{67}},
  \bibinfo{pages}{084023} (\bibinfo{year}{2003}), \eprint{gr-qc/0206072}.

\bibitem[{\citenamefont{Anninos et~al.}(1996)\citenamefont{Anninos, Mass{\'o},
  Seidel, and Suen}}]{Anninos96c}
\bibinfo{author}{\bibfnamefont{P.}~\bibnamefont{Anninos}},
  \bibinfo{author}{\bibfnamefont{J.}~\bibnamefont{Mass{\'o}}},
  \bibinfo{author}{\bibfnamefont{E.}~\bibnamefont{Seidel}}, \bibnamefont{and}
  \bibinfo{author}{\bibfnamefont{W.-M.} \bibnamefont{Suen}},
  \bibinfo{journal}{Physics World} \textbf{\bibinfo{volume}{9}},
  \bibinfo{pages}{43} (\bibinfo{year}{1996}).

\bibitem[{\citenamefont{Baker et~al.}(2000)\citenamefont{Baker, Br\"ugmann,
  Campanelli, and Lousto}}]{Baker00b}
\bibinfo{author}{\bibfnamefont{J.}~\bibnamefont{Baker}},
  \bibinfo{author}{\bibfnamefont{B.}~\bibnamefont{Br\"ugmann}},
  \bibinfo{author}{\bibfnamefont{M.}~\bibnamefont{Campanelli}},
  \bibnamefont{and} \bibinfo{author}{\bibfnamefont{C.~O.}
  \bibnamefont{Lousto}}, \bibinfo{journal}{Class. Quantum Grav.}
  \textbf{\bibinfo{volume}{17}}, \bibinfo{pages}{L149} (\bibinfo{year}{2000}),
  \eprint{gr-qc/0003027}.

\bibitem[{\citenamefont{Baker et~al.}(2001)\citenamefont{Baker, Br{\"u}gmann,
  Campanelli, Lousto, and Takahashi}}]{Baker:2001nu}
\bibinfo{author}{\bibfnamefont{J.}~\bibnamefont{Baker}},
  \bibinfo{author}{\bibfnamefont{B.}~\bibnamefont{Br{\"u}gmann}},
  \bibinfo{author}{\bibfnamefont{M.}~\bibnamefont{Campanelli}},
  \bibinfo{author}{\bibfnamefont{C.~O.} \bibnamefont{Lousto}},
  \bibnamefont{and}
  \bibinfo{author}{\bibfnamefont{R.}~\bibnamefont{Takahashi}},
  \bibinfo{journal}{Phys. Rev. Lett.} \textbf{\bibinfo{volume}{87}},
  \bibinfo{pages}{121103} (\bibinfo{year}{2001}),
  \eprint[http://arXiv.org/abs]{gr-qc/0102037}.

\bibitem[{\citenamefont{Br{\"u}gmann}(1999)}]{Bruegmann97}
\bibinfo{author}{\bibfnamefont{B.}~\bibnamefont{Br{\"u}gmann}},
  \bibinfo{journal}{Int. J. Mod. Phys. D} \textbf{\bibinfo{volume}{8}},
  \bibinfo{pages}{85} (\bibinfo{year}{1999}).

\bibitem[{\citenamefont{Brandt et~al.}(2000)\citenamefont{Brandt, Correll,
  G\'{o}mez, Huq, Laguna, Lehner, Marronetti, Matzner, Neilsen, Pullin
  et~al.}}]{Brandt00}
\bibinfo{author}{\bibfnamefont{S.}~\bibnamefont{Brandt}},
  \bibinfo{author}{\bibfnamefont{R.}~\bibnamefont{Correll}},
  \bibinfo{author}{\bibfnamefont{R.}~\bibnamefont{G\'{o}mez}},
  \bibinfo{author}{\bibfnamefont{M.~F.} \bibnamefont{Huq}},
  \bibinfo{author}{\bibfnamefont{P.}~\bibnamefont{Laguna}},
  \bibinfo{author}{\bibfnamefont{L.}~\bibnamefont{Lehner}},
  \bibinfo{author}{\bibfnamefont{P.}~\bibnamefont{Marronetti}},
  \bibinfo{author}{\bibfnamefont{R.~A.} \bibnamefont{Matzner}},
  \bibinfo{author}{\bibfnamefont{D.}~\bibnamefont{Neilsen}},
  \bibinfo{author}{\bibfnamefont{J.}~\bibnamefont{Pullin}},
  \bibnamefont{et~al.}, \bibinfo{journal}{Phys. Rev. Lett.}
  \textbf{\bibinfo{volume}{85}}, \bibinfo{pages}{5496} (\bibinfo{year}{2000}).

\bibitem[{\citenamefont{Alcubierre
  et~al.}(2001{\natexlab{a}})\citenamefont{Alcubierre, Benger, Br\"ugmann,
  Lanfermann, Nerger, Seidel, and Takahashi}}]{Alcubierre00b}
\bibinfo{author}{\bibfnamefont{M.}~\bibnamefont{Alcubierre}},
  \bibinfo{author}{\bibfnamefont{W.}~\bibnamefont{Benger}},
  \bibinfo{author}{\bibfnamefont{B.}~\bibnamefont{Br\"ugmann}},
  \bibinfo{author}{\bibfnamefont{G.}~\bibnamefont{Lanfermann}},
  \bibinfo{author}{\bibfnamefont{L.}~\bibnamefont{Nerger}},
  \bibinfo{author}{\bibfnamefont{E.}~\bibnamefont{Seidel}}, \bibnamefont{and}
  \bibinfo{author}{\bibfnamefont{R.}~\bibnamefont{Takahashi}},
  \bibinfo{journal}{Phys. Rev. Lett.} \textbf{\bibinfo{volume}{87}},
  \bibinfo{pages}{271103} (\bibinfo{year}{2001}{\natexlab{a}}),
  \eprint{gr-qc/0012079}.

\bibitem[{\citenamefont{Seidel}(2000)}]{Seidel99b}
\bibinfo{author}{\bibfnamefont{E.}~\bibnamefont{Seidel}},
  \bibinfo{journal}{Prog. Theor. Phys. Suppl.} \textbf{\bibinfo{volume}{136}},
  \bibinfo{pages}{87} (\bibinfo{year}{2000}).

\bibitem[{\citenamefont{Br\"ugmann et~al.}(2004)\citenamefont{Br\"ugmann,
  Tichy, and Jansen}}]{Bruegmann:2003aw}
\bibinfo{author}{\bibfnamefont{B.}~\bibnamefont{Br\"ugmann}},
  \bibinfo{author}{\bibfnamefont{W.}~\bibnamefont{Tichy}}, \bibnamefont{and}
  \bibinfo{author}{\bibfnamefont{N.}~\bibnamefont{Jansen}},
  \bibinfo{journal}{Phys. Rev. Lett.} \textbf{\bibinfo{volume}{92}},
  \bibinfo{pages}{211101} (\bibinfo{year}{2004}), \eprint{gr-qc/0312112}.

\bibitem[{\citenamefont{Cook}(1994)}]{Cook94}
\bibinfo{author}{\bibfnamefont{G.~B.} \bibnamefont{Cook}},
  \bibinfo{journal}{Phys. Rev. D} \textbf{\bibinfo{volume}{50}},
  \bibinfo{pages}{5025} (\bibinfo{year}{1994}).

\bibitem[{\citenamefont{Baumgarte}(2000)}]{Baumgarte00a}
\bibinfo{author}{\bibfnamefont{T.~W.} \bibnamefont{Baumgarte}},
  \bibinfo{journal}{Phys. Rev. D} \textbf{\bibinfo{volume}{62}},
  \bibinfo{pages}{024018} (\bibinfo{year}{2000}), \eprint{gr-qc/0004050}.

\bibitem[{\citenamefont{Tichy et~al.}(2003)\citenamefont{Tichy, Br\"ugmann, and
  Laguna}}]{Tichy03a}
\bibinfo{author}{\bibfnamefont{W.}~\bibnamefont{Tichy}},
  \bibinfo{author}{\bibfnamefont{B.}~\bibnamefont{Br\"ugmann}},
  \bibnamefont{and} \bibinfo{author}{\bibfnamefont{P.}~\bibnamefont{Laguna}},
  \bibinfo{journal}{Phys. Rev. D} \textbf{\bibinfo{volume}{68}},
  \bibinfo{pages}{064008} (\bibinfo{year}{2003}), \eprint{gr-qc/0306020}.

\bibitem[{\citenamefont{Tichy and Br{\"u}gmann}(2004)}]{Tichy:2003qi}
\bibinfo{author}{\bibfnamefont{W.}~\bibnamefont{Tichy}} \bibnamefont{and}
  \bibinfo{author}{\bibfnamefont{B.}~\bibnamefont{Br{\"u}gmann}},
  \bibinfo{journal}{Phys. Rev. D} \textbf{\bibinfo{volume}{69}},
  \bibinfo{pages}{024006} (\bibinfo{year}{2004}), \eprint{gr-qc/0307027}.

\bibitem[{\citenamefont{Gourgoulhon et~al.}(2002)\citenamefont{Gourgoulhon,
  Grandcl\'{e}ment, and Bonazzola}}]{Gourgoulhon02}
\bibinfo{author}{\bibfnamefont{E.}~\bibnamefont{Gourgoulhon}},
  \bibinfo{author}{\bibfnamefont{P.}~\bibnamefont{Grandcl\'{e}ment}},
  \bibnamefont{and}
  \bibinfo{author}{\bibfnamefont{S.}~\bibnamefont{Bonazzola}},
  \bibinfo{journal}{Phys. Rev. D} \textbf{\bibinfo{volume}{65}},
  \bibinfo{pages}{044020} (\bibinfo{year}{2002}), \eprint{gr-qc/0106015}.

\bibitem[{\citenamefont{Grandcl\'{e}ment
  et~al.}(2002)\citenamefont{Grandcl\'{e}ment, Gourgoulhon, and
  Bonazzola}}]{Grandclement02}
\bibinfo{author}{\bibfnamefont{P.}~\bibnamefont{Grandcl\'{e}ment}},
  \bibinfo{author}{\bibfnamefont{E.}~\bibnamefont{Gourgoulhon}},
  \bibnamefont{and}
  \bibinfo{author}{\bibfnamefont{S.}~\bibnamefont{Bonazzola}},
  \bibinfo{journal}{Phys. Rev. D} \textbf{\bibinfo{volume}{65}},
  \bibinfo{pages}{044021} (\bibinfo{year}{2002}), \eprint{gr-qc/0106016}.

\bibitem[{\citenamefont{Cook}(2002)}]{Cook:2001wi}
\bibinfo{author}{\bibfnamefont{G.~B.} \bibnamefont{Cook}},
  \bibinfo{journal}{Phys. Rev. D} \textbf{\bibinfo{volume}{65}},
  \bibinfo{pages}{084003} (\bibinfo{year}{2002}),
  \eprint[http://arXiv.org/abs]{gr-qc/0108076}.

\bibitem[{\citenamefont{Damour et~al.}(2002)\citenamefont{Damour, Gourgoulhon,
  and Grandcl\'{e}ment}}]{Damour:2002qh}
\bibinfo{author}{\bibfnamefont{T.}~\bibnamefont{Damour}},
  \bibinfo{author}{\bibfnamefont{E.}~\bibnamefont{Gourgoulhon}},
  \bibnamefont{and}
  \bibinfo{author}{\bibfnamefont{P.}~\bibnamefont{Grandcl\'{e}ment}},
  \bibinfo{journal}{Phys. Rev. D} \textbf{\bibinfo{volume}{66}},
  \bibinfo{pages}{024007} (\bibinfo{year}{2002}),
  \eprint[http://arXiv.org/abs]{gr-qc/0204011}.

\bibitem[{\citenamefont{Nakamura et~al.}(1987)\citenamefont{Nakamura, Oohara,
  and Kojima}}]{Nakamura87}
\bibinfo{author}{\bibfnamefont{T.}~\bibnamefont{Nakamura}},
  \bibinfo{author}{\bibfnamefont{K.}~\bibnamefont{Oohara}}, \bibnamefont{and}
  \bibinfo{author}{\bibfnamefont{Y.}~\bibnamefont{Kojima}},
  \bibinfo{journal}{Prog. Theor. Phys. Suppl.} \textbf{\bibinfo{volume}{90}},
  \bibinfo{pages}{1} (\bibinfo{year}{1987}).

\bibitem[{\citenamefont{Shibata and Nakamura}(1995)}]{Shibata95}
\bibinfo{author}{\bibfnamefont{M.}~\bibnamefont{Shibata}} \bibnamefont{and}
  \bibinfo{author}{\bibfnamefont{T.}~\bibnamefont{Nakamura}},
  \bibinfo{journal}{Phys. Rev. D} \textbf{\bibinfo{volume}{52}},
  \bibinfo{pages}{5428} (\bibinfo{year}{1995}).

\bibitem[{\citenamefont{Baumgarte and Shapiro}(1999)}]{Baumgarte99}
\bibinfo{author}{\bibfnamefont{T.~W.} \bibnamefont{Baumgarte}}
  \bibnamefont{and} \bibinfo{author}{\bibfnamefont{S.~L.}
  \bibnamefont{Shapiro}}, \bibinfo{journal}{Phys. Rev. D}
  \textbf{\bibinfo{volume}{59}}, \bibinfo{pages}{024007}
  (\bibinfo{year}{1999}), \eprint{gr-qc/9810065}.

\bibitem[{\citenamefont{Alcubierre
  et~al.}(2000{\natexlab{a}})\citenamefont{Alcubierre, Br\"{u}gmann,
  Dramlitsch, Font, Papadopoulos, Seidel, Stergioulas, and
  Takahashi}}]{Alcubierre99d}
\bibinfo{author}{\bibfnamefont{M.}~\bibnamefont{Alcubierre}},
  \bibinfo{author}{\bibfnamefont{B.}~\bibnamefont{Br\"{u}gmann}},
  \bibinfo{author}{\bibfnamefont{T.}~\bibnamefont{Dramlitsch}},
  \bibinfo{author}{\bibfnamefont{J.~A.} \bibnamefont{Font}},
  \bibinfo{author}{\bibfnamefont{P.}~\bibnamefont{Papadopoulos}},
  \bibinfo{author}{\bibfnamefont{E.}~\bibnamefont{Seidel}},
  \bibinfo{author}{\bibfnamefont{N.}~\bibnamefont{Stergioulas}},
  \bibnamefont{and}
  \bibinfo{author}{\bibfnamefont{R.}~\bibnamefont{Takahashi}},
  \bibinfo{journal}{Phys. Rev. D} \textbf{\bibinfo{volume}{62}},
  \bibinfo{pages}{044034} (\bibinfo{year}{2000}{\natexlab{a}}),
  \eprint{gr-qc/0003071}.

\bibitem[{\citenamefont{Alcubierre
  et~al.}(2001{\natexlab{b}})\citenamefont{Alcubierre, Br\"ugmann, Pollney,
  Seidel, and Takahashi}}]{Alcubierre01a}
\bibinfo{author}{\bibfnamefont{M.}~\bibnamefont{Alcubierre}},
  \bibinfo{author}{\bibfnamefont{B.}~\bibnamefont{Br\"ugmann}},
  \bibinfo{author}{\bibfnamefont{D.}~\bibnamefont{Pollney}},
  \bibinfo{author}{\bibfnamefont{E.}~\bibnamefont{Seidel}}, \bibnamefont{and}
  \bibinfo{author}{\bibfnamefont{R.}~\bibnamefont{Takahashi}},
  \bibinfo{journal}{Phys. Rev. D} \textbf{\bibinfo{volume}{64}},
  \bibinfo{pages}{061501(R)} (\bibinfo{year}{2001}{\natexlab{b}}),
  \eprint{gr-qc/0104020}.

\bibitem[{\citenamefont{Alcubierre et~al.}(2004)\citenamefont{Alcubierre,
  Br\"ugmann, Diener, Herrmann, Pollney, Seidel, and
  Takahashi}}]{Alcubierre2003:BBH0-excision}
\bibinfo{author}{\bibfnamefont{M.}~\bibnamefont{Alcubierre}},
  \bibinfo{author}{\bibfnamefont{B.}~\bibnamefont{Br\"ugmann}},
  \bibinfo{author}{\bibfnamefont{P.}~\bibnamefont{Diener}},
  \bibinfo{author}{\bibfnamefont{F.}~\bibnamefont{Herrmann}},
  \bibinfo{author}{\bibfnamefont{D.}~\bibnamefont{Pollney}},
  \bibinfo{author}{\bibfnamefont{E.}~\bibnamefont{Seidel}}, \bibnamefont{and}
  \bibinfo{author}{\bibfnamefont{R.}~\bibnamefont{Takahashi}},
  \bibinfo{journal}{submitted to Phys. Rev. D}  (\bibinfo{year}{2004}),
  \eprint{gr-qc/0411137}.

\bibitem[{\citenamefont{Alcubierre et~al.}(2005)\citenamefont{Alcubierre,
  Diener, Guzm\'an, Hawley, Koppitz, Pollney, and
  Seidel}}]{Alcubierre2003:co-rotating-shift}
\bibinfo{author}{\bibfnamefont{M.}~\bibnamefont{Alcubierre}},
  \bibinfo{author}{\bibfnamefont{P.}~\bibnamefont{Diener}},
  \bibinfo{author}{\bibfnamefont{F.~S.} \bibnamefont{Guzm\'an}},
  \bibinfo{author}{\bibfnamefont{S.}~\bibnamefont{Hawley}},
  \bibinfo{author}{\bibfnamefont{M.}~\bibnamefont{Koppitz}},
  \bibinfo{author}{\bibfnamefont{D.}~\bibnamefont{Pollney}}, \bibnamefont{and}
  \bibinfo{author}{\bibfnamefont{E.}~\bibnamefont{Seidel}}
  (\bibinfo{year}{2005}), \bibinfo{note}{in preparation}.

\bibitem[{\citenamefont{Baker et~al.}(2002{\natexlab{a}})\citenamefont{Baker,
  Campanelli, Lousto, and Takahashi}}]{Baker:2002qf}
\bibinfo{author}{\bibfnamefont{J.}~\bibnamefont{Baker}},
  \bibinfo{author}{\bibfnamefont{M.}~\bibnamefont{Campanelli}},
  \bibinfo{author}{\bibfnamefont{C.~O.} \bibnamefont{Lousto}},
  \bibnamefont{and}
  \bibinfo{author}{\bibfnamefont{R.}~\bibnamefont{Takahashi}},
  \bibinfo{journal}{Phys. Rev. D} \textbf{\bibinfo{volume}{65}},
  \bibinfo{pages}{124012} (\bibinfo{year}{2002}{\natexlab{a}}),
  \eprint[http://arXiv.org/abs]{astro-ph/0202469}.

\bibitem[{\citenamefont{Abrahams et~al.}(1992)\citenamefont{Abrahams,
  Bernstein, Hobill, Seidel, and Smarr}}]{Abrahams92a}
\bibinfo{author}{\bibfnamefont{A.}~\bibnamefont{Abrahams}},
  \bibinfo{author}{\bibfnamefont{D.}~\bibnamefont{Bernstein}},
  \bibinfo{author}{\bibfnamefont{D.}~\bibnamefont{Hobill}},
  \bibinfo{author}{\bibfnamefont{E.}~\bibnamefont{Seidel}}, \bibnamefont{and}
  \bibinfo{author}{\bibfnamefont{L.}~\bibnamefont{Smarr}},
  \bibinfo{journal}{Phys. Rev. D} \textbf{\bibinfo{volume}{45}},
  \bibinfo{pages}{3544} (\bibinfo{year}{1992}).

\bibitem[{\citenamefont{Abrahams et~al.}(1995)\citenamefont{Abrahams, Shapiro,
  and {T}eukolsky}}]{Abrahams95d}
\bibinfo{author}{\bibfnamefont{A.~M.} \bibnamefont{Abrahams}},
  \bibinfo{author}{\bibfnamefont{S.~L.} \bibnamefont{Shapiro}},
  \bibnamefont{and} \bibinfo{author}{\bibfnamefont{S.~A.}
  \bibnamefont{{T}eukolsky}}, \bibinfo{journal}{Phys. Rev. D}
  \textbf{\bibinfo{volume}{51}}, \bibinfo{pages}{4295} (\bibinfo{year}{1995}),
  \eprint{gr-qc/9408036}.

\bibitem[{\citenamefont{Anninos
  et~al.}(1995{\natexlab{a}})\citenamefont{Anninos, Bernstein, Brandt, Libson,
  Mass{\'o}, Seidel, Smarr, Suen, and Walker}}]{Anninos94f}
\bibinfo{author}{\bibfnamefont{P.}~\bibnamefont{Anninos}},
  \bibinfo{author}{\bibfnamefont{D.}~\bibnamefont{Bernstein}},
  \bibinfo{author}{\bibfnamefont{S.}~\bibnamefont{Brandt}},
  \bibinfo{author}{\bibfnamefont{J.}~\bibnamefont{Libson}},
  \bibinfo{author}{\bibfnamefont{J.}~\bibnamefont{Mass{\'o}}},
  \bibinfo{author}{\bibfnamefont{E.}~\bibnamefont{Seidel}},
  \bibinfo{author}{\bibfnamefont{L.}~\bibnamefont{Smarr}},
  \bibinfo{author}{\bibfnamefont{W.-M.} \bibnamefont{Suen}}, \bibnamefont{and}
  \bibinfo{author}{\bibfnamefont{P.}~\bibnamefont{Walker}},
  \bibinfo{journal}{Phys. Rev. Lett.} \textbf{\bibinfo{volume}{74}},
  \bibinfo{pages}{630} (\bibinfo{year}{1995}{\natexlab{a}}).

\bibitem[{\citenamefont{Anninos
  et~al.}(1995{\natexlab{b}})\citenamefont{Anninos, Bernstein, Brandt, Hobill,
  Seidel, and Smarr}}]{Anninos95c}
\bibinfo{author}{\bibfnamefont{P.}~\bibnamefont{Anninos}},
  \bibinfo{author}{\bibfnamefont{D.}~\bibnamefont{Bernstein}},
  \bibinfo{author}{\bibfnamefont{S.}~\bibnamefont{Brandt}},
  \bibinfo{author}{\bibfnamefont{D.}~\bibnamefont{Hobill}},
  \bibinfo{author}{\bibfnamefont{E.}~\bibnamefont{Seidel}}, \bibnamefont{and}
  \bibinfo{author}{\bibfnamefont{L.}~\bibnamefont{Smarr}},
  \bibinfo{journal}{Australian Journal of Physics}
  \textbf{\bibinfo{volume}{48}}, \bibinfo{pages}{1027}
  (\bibinfo{year}{1995}{\natexlab{b}}).

\bibitem[{\citenamefont{Brandt and Seidel}(1995)}]{Brandt94c}
\bibinfo{author}{\bibfnamefont{S.~R.} \bibnamefont{Brandt}} \bibnamefont{and}
  \bibinfo{author}{\bibfnamefont{E.}~\bibnamefont{Seidel}},
  \bibinfo{journal}{Phys. Rev. D} \textbf{\bibinfo{volume}{52}},
  \bibinfo{pages}{870} (\bibinfo{year}{1995}).

\bibitem[{\citenamefont{Libson et~al.}(1996)\citenamefont{Libson, Mass{\'o},
  Seidel, Suen, and Walker}}]{Libson94a}
\bibinfo{author}{\bibfnamefont{J.}~\bibnamefont{Libson}},
  \bibinfo{author}{\bibfnamefont{J.}~\bibnamefont{Mass{\'o}}},
  \bibinfo{author}{\bibfnamefont{E.}~\bibnamefont{Seidel}},
  \bibinfo{author}{\bibfnamefont{W.-M.} \bibnamefont{Suen}}, \bibnamefont{and}
  \bibinfo{author}{\bibfnamefont{P.}~\bibnamefont{Walker}},
  \bibinfo{journal}{Phys. Rev. D} \textbf{\bibinfo{volume}{53}},
  \bibinfo{pages}{4335} (\bibinfo{year}{1996}).

\bibitem[{\citenamefont{Mass{\'o} et~al.}(1999)\citenamefont{Mass{\'o}, Seidel,
  Suen, and Walker}}]{Masso95a}
\bibinfo{author}{\bibfnamefont{J.}~\bibnamefont{Mass{\'o}}},
  \bibinfo{author}{\bibfnamefont{E.}~\bibnamefont{Seidel}},
  \bibinfo{author}{\bibfnamefont{W.-M.} \bibnamefont{Suen}}, \bibnamefont{and}
  \bibinfo{author}{\bibfnamefont{P.}~\bibnamefont{Walker}},
  \bibinfo{journal}{Phys. Rev. D} \textbf{\bibinfo{volume}{59}},
  \bibinfo{pages}{064015} (\bibinfo{year}{1999}),
  \bibinfo{note}{gr-qc/9804059}.

\bibitem[{\citenamefont{Br{\"u}gmann}(1996)}]{Bruegmann96}
\bibinfo{author}{\bibfnamefont{B.}~\bibnamefont{Br{\"u}gmann}},
  \bibinfo{journal}{Phys. Rev. D} \textbf{\bibinfo{volume}{54}},
  \bibinfo{pages}{7361} (\bibinfo{year}{1996}), \eprint{gr-qc/9608050}.

\bibitem[{\citenamefont{Br{\"u}gmann}(2000)}]{Bruegmann99b}
\bibinfo{author}{\bibfnamefont{B.}~\bibnamefont{Br{\"u}gmann}},
  \bibinfo{journal}{Ann. Phys. (Leipzig)} \textbf{\bibinfo{volume}{9}},
  \bibinfo{pages}{227} (\bibinfo{year}{2000}), \bibinfo{note}{gr-qc/9912009}.

\bibitem[{\citenamefont{Dreyer et~al.}(2002)\citenamefont{Dreyer, Krishnan,
  Shoemaker, and Schnetter}}]{Dreyer-etal-2002-isolated-horizons}
\bibinfo{author}{\bibfnamefont{O.}~\bibnamefont{Dreyer}},
  \bibinfo{author}{\bibfnamefont{B.}~\bibnamefont{Krishnan}},
  \bibinfo{author}{\bibfnamefont{D.}~\bibnamefont{Shoemaker}},
  \bibnamefont{and}
  \bibinfo{author}{\bibfnamefont{E.}~\bibnamefont{Schnetter}},
  \bibinfo{journal}{Phys. Rev. D} \textbf{\bibinfo{volume}{67}},
  \bibinfo{pages}{024018} (\bibinfo{year}{2002}), \eprint{gr-qc/0206008}.

\bibitem[{\citenamefont{Ashtekar et~al.}(1999)\citenamefont{Ashtekar, Beetle,
  and Fairhurst}}]{Ashtekar98a}
\bibinfo{author}{\bibfnamefont{A.}~\bibnamefont{Ashtekar}},
  \bibinfo{author}{\bibfnamefont{C.}~\bibnamefont{Beetle}}, \bibnamefont{and}
  \bibinfo{author}{\bibfnamefont{S.}~\bibnamefont{Fairhurst}},
  \bibinfo{journal}{Class. Quantum Grav.} \textbf{\bibinfo{volume}{16}},
  \bibinfo{pages}{L1} (\bibinfo{year}{1999}), \eprint{gr-qc/9812065}.

\bibitem[{\citenamefont{Ashtekar and Krishnan}(2004)}]{Ashtekar:2004cn}
\bibinfo{author}{\bibfnamefont{A.}~\bibnamefont{Ashtekar}} \bibnamefont{and}
  \bibinfo{author}{\bibfnamefont{B.}~\bibnamefont{Krishnan}},
  \bibinfo{journal}{Living Rev. Rel.} \textbf{\bibinfo{volume}{7}},
  \bibinfo{pages}{10} (\bibinfo{year}{2004}), \eprint{gr-qc/0407042}.

\bibitem[{\citenamefont{Brandt and Br{\"u}gmann}(1997)}]{Brandt97b}
\bibinfo{author}{\bibfnamefont{S.}~\bibnamefont{Brandt}} \bibnamefont{and}
  \bibinfo{author}{\bibfnamefont{B.}~\bibnamefont{Br{\"u}gmann}},
  \bibinfo{journal}{Phys. Rev. Lett.} \textbf{\bibinfo{volume}{78}},
  \bibinfo{pages}{3606} (\bibinfo{year}{1997}), \eprint{gr-qc/9703066}.

\bibitem[{\citenamefont{Thornburg}(2004)}]{Thornburg2003:AH-finding}
\bibinfo{author}{\bibfnamefont{J.}~\bibnamefont{Thornburg}},
  \bibinfo{journal}{Class. Quantum Grav.} \textbf{\bibinfo{volume}{21}},
  \bibinfo{pages}{743} (\bibinfo{year}{2004}), \eprint{gr-qc/0306056},
  \urlprefix\url{http://stacks.iop.org/0264-9381/21/743}.

\bibitem[{\citenamefont{Gundlach}(1998)}]{Gundlach97a}
\bibinfo{author}{\bibfnamefont{C.}~\bibnamefont{Gundlach}},
  \bibinfo{journal}{Phys. Rev. D} \textbf{\bibinfo{volume}{57}},
  \bibinfo{pages}{863} (\bibinfo{year}{1998}), \bibinfo{note}{gr-qc/9707050}.

\bibitem[{\citenamefont{Alcubierre
  et~al.}(2000{\natexlab{b}})\citenamefont{Alcubierre, Brandt, Br{\"u}gmann,
  Gundlach, Mass{\'o}, Seidel, and Walker}}]{Alcubierre98b}
\bibinfo{author}{\bibfnamefont{M.}~\bibnamefont{Alcubierre}},
  \bibinfo{author}{\bibfnamefont{S.}~\bibnamefont{Brandt}},
  \bibinfo{author}{\bibfnamefont{B.}~\bibnamefont{Br{\"u}gmann}},
  \bibinfo{author}{\bibfnamefont{C.}~\bibnamefont{Gundlach}},
  \bibinfo{author}{\bibfnamefont{J.}~\bibnamefont{Mass{\'o}}},
  \bibinfo{author}{\bibfnamefont{E.}~\bibnamefont{Seidel}}, \bibnamefont{and}
  \bibinfo{author}{\bibfnamefont{P.}~\bibnamefont{Walker}},
  \bibinfo{journal}{Class. Quantum Grav.} \textbf{\bibinfo{volume}{17}},
  \bibinfo{pages}{2159} (\bibinfo{year}{2000}{\natexlab{b}}),
  \eprint{gr-qc/9809004}.

\bibitem[{\citenamefont{Diener}(2003)}]{Diener03a}
\bibinfo{author}{\bibfnamefont{P.}~\bibnamefont{Diener}},
  \bibinfo{journal}{Class. Quantum Grav.} \textbf{\bibinfo{volume}{20}},
  \bibinfo{pages}{4901} (\bibinfo{year}{2003}), \eprint{gr-qc/0305039}.

\bibitem[{\citenamefont{Alcubierre and Br\"ugmann}(2001)}]{Alcubierre00a}
\bibinfo{author}{\bibfnamefont{M.}~\bibnamefont{Alcubierre}} \bibnamefont{and}
  \bibinfo{author}{\bibfnamefont{B.}~\bibnamefont{Br\"ugmann}},
  \bibinfo{journal}{Phys. Rev. D} \textbf{\bibinfo{volume}{63}},
  \bibinfo{pages}{104006} (\bibinfo{year}{2001}), \eprint{gr-qc/0008067}.

\bibitem[{\citenamefont{Anninos
  et~al.}(1995{\natexlab{c}})\citenamefont{Anninos, Hobill, Seidel, Smarr, and
  Suen}}]{Anninos94b}
\bibinfo{author}{\bibfnamefont{P.}~\bibnamefont{Anninos}},
  \bibinfo{author}{\bibfnamefont{D.}~\bibnamefont{Hobill}},
  \bibinfo{author}{\bibfnamefont{E.}~\bibnamefont{Seidel}},
  \bibinfo{author}{\bibfnamefont{L.}~\bibnamefont{Smarr}}, \bibnamefont{and}
  \bibinfo{author}{\bibfnamefont{W.-M.} \bibnamefont{Suen}},
  \bibinfo{journal}{Phys. Rev. D} \textbf{\bibinfo{volume}{52}},
  \bibinfo{pages}{2044} (\bibinfo{year}{1995}{\natexlab{c}}).

\bibitem[{\citenamefont{Smarr}(1973{\natexlab{a}})}]{Smarr73a}
\bibinfo{author}{\bibfnamefont{L.~L.} \bibnamefont{Smarr}},
  \bibinfo{journal}{Phys. Rev. Lett.} \textbf{\bibinfo{volume}{30}},
  \bibinfo{pages}{71} (\bibinfo{year}{1973}{\natexlab{a}}).

\bibitem[{\citenamefont{Christodoulou}(1970)}]{Christodoulou70}
\bibinfo{author}{\bibfnamefont{D.}~\bibnamefont{Christodoulou}},
  \bibinfo{journal}{Phys. Rev. Lett.} \textbf{\bibinfo{volume}{25}},
  \bibinfo{pages}{1596} (\bibinfo{year}{1970}).

\bibitem[{\citenamefont{Matzner et~al.}(1995)\citenamefont{Matzner, Seidel,
  Shapiro, Smarr, Suen, Teukolsky, and Winicour}}]{Matzner95a}
\bibinfo{author}{\bibfnamefont{R.~A.} \bibnamefont{Matzner}},
  \bibinfo{author}{\bibfnamefont{E.}~\bibnamefont{Seidel}},
  \bibinfo{author}{\bibfnamefont{S.}~\bibnamefont{Shapiro}},
  \bibinfo{author}{\bibfnamefont{L.}~\bibnamefont{Smarr}},
  \bibinfo{author}{\bibfnamefont{W.-M.} \bibnamefont{Suen}},
  \bibinfo{author}{\bibfnamefont{S.}~\bibnamefont{Teukolsky}},
  \bibnamefont{and} \bibinfo{author}{\bibfnamefont{J.}~\bibnamefont{Winicour}},
  \bibinfo{journal}{Science} \textbf{\bibinfo{volume}{270}},
  \bibinfo{pages}{941} (\bibinfo{year}{1995}).

\bibitem[{\citenamefont{Caveny et~al.}(2003)\citenamefont{Caveny, Anderson, and
  Matzner}}]{Caveny-Anderson-Matzner-2003a}
\bibinfo{author}{\bibfnamefont{S.~A.} \bibnamefont{Caveny}},
  \bibinfo{author}{\bibfnamefont{M.}~\bibnamefont{Anderson}}, \bibnamefont{and}
  \bibinfo{author}{\bibfnamefont{R.~A.} \bibnamefont{Matzner}},
  \bibinfo{journal}{Phys. Rev. D} \textbf{\bibinfo{volume}{68}},
  \bibinfo{pages}{104009} (\bibinfo{year}{2003}), \eprint{gr-qc/0303099}.

\bibitem[{\citenamefont{Smarr}(1973{\natexlab{b}})}]{Smarr73b}
\bibinfo{author}{\bibfnamefont{L.~L.} \bibnamefont{Smarr}},
  \bibinfo{journal}{Phys. Rev. D} \textbf{\bibinfo{volume}{7}},
  \bibinfo{pages}{289} (\bibinfo{year}{1973}{\natexlab{b}}).

\bibitem[{\citenamefont{Leaver}(1986)}]{Leaver86}
\bibinfo{author}{\bibfnamefont{E.~W.} \bibnamefont{Leaver}},
  \bibinfo{journal}{Proc. R. Soc. London, Series A}
  \textbf{\bibinfo{volume}{402}}, \bibinfo{pages}{285} (\bibinfo{year}{1986}).

\bibitem[{\citenamefont{Seidel and Iyer}(1990)}]{Seidel90a}
\bibinfo{author}{\bibfnamefont{E.}~\bibnamefont{Seidel}} \bibnamefont{and}
  \bibinfo{author}{\bibfnamefont{S.}~\bibnamefont{Iyer}},
  \bibinfo{journal}{Phys. Rev. D} \textbf{\bibinfo{volume}{41}},
  \bibinfo{pages}{374} (\bibinfo{year}{1990}).

\bibitem[{\citenamefont{Ashtekar et~al.}(2000)\citenamefont{Ashtekar, Beetle,
  and Fairhurst}}]{Ashtekar99a}
\bibinfo{author}{\bibfnamefont{A.}~\bibnamefont{Ashtekar}},
  \bibinfo{author}{\bibfnamefont{C.}~\bibnamefont{Beetle}}, \bibnamefont{and}
  \bibinfo{author}{\bibfnamefont{S.}~\bibnamefont{Fairhurst}},
  \bibinfo{journal}{Class. Quantum Grav.} \textbf{\bibinfo{volume}{17}},
  \bibinfo{pages}{253} (\bibinfo{year}{2000}), \eprint{gr-qc/9907068}.

\bibitem[{\citenamefont{Ashtekar and
  Krishnan}(2002)}]{Ashtekar-etal-2002-dynamical-horizons}
\bibinfo{author}{\bibfnamefont{A.}~\bibnamefont{Ashtekar}} \bibnamefont{and}
  \bibinfo{author}{\bibfnamefont{B.}~\bibnamefont{Krishnan}},
  \bibinfo{journal}{Phys. Rev. Lett.} \textbf{\bibinfo{volume}{89}},
  \bibinfo{pages}{261101} (\bibinfo{year}{2002}), \eprint{gr-qc/0207080}.

\bibitem[{\citenamefont{Diener et~al.}()\citenamefont{Diener, Takahashi,
  Pollney, and Seidel}}]{Diener2004:dist-bh-rotating}
\bibinfo{author}{\bibfnamefont{P.}~\bibnamefont{Diener}},
  \bibinfo{author}{\bibfnamefont{R.}~\bibnamefont{Takahashi}},
  \bibinfo{author}{\bibfnamefont{D.}~\bibnamefont{Pollney}}, \bibnamefont{and}
  \bibinfo{author}{\bibfnamefont{E.}~\bibnamefont{Seidel}}, \bibinfo{note}{in
  preparation}.

\bibitem[{\citenamefont{Baker et~al.}(2002{\natexlab{b}})\citenamefont{Baker,
  Campanelli, and Lousto}}]{Baker:2001sf}
\bibinfo{author}{\bibfnamefont{J.}~\bibnamefont{Baker}},
  \bibinfo{author}{\bibfnamefont{M.}~\bibnamefont{Campanelli}},
  \bibnamefont{and} \bibinfo{author}{\bibfnamefont{C.~O.}
  \bibnamefont{Lousto}}, \bibinfo{journal}{Phys. Rev. D}
  \textbf{\bibinfo{volume}{65}}, \bibinfo{pages}{044001}
  (\bibinfo{year}{2002}{\natexlab{b}}),
  \eprint[http://arXiv.org/abs]{gr-qc/0104063}.

\bibitem[{\citenamefont{Miller}(2004)}]{Miller03c}
\bibinfo{author}{\bibfnamefont{M.}~\bibnamefont{Miller}},
  \bibinfo{journal}{Phys. Rev. D} \textbf{\bibinfo{volume}{69}},
  \bibinfo{pages}{124013} (\bibinfo{year}{2004}),
  \bibinfo{note}{gr-qc/0305024}.

\bibitem[{\citenamefont{Miller et~al.}(2004)\citenamefont{Miller, Gressman, and
  Suen}}]{Miller:2003vc}
\bibinfo{author}{\bibfnamefont{M.}~\bibnamefont{Miller}},
  \bibinfo{author}{\bibfnamefont{P.}~\bibnamefont{Gressman}}, \bibnamefont{and}
  \bibinfo{author}{\bibfnamefont{W.-M.} \bibnamefont{Suen}},
  \bibinfo{journal}{Phys. Rev. D} \textbf{\bibinfo{volume}{69}},
  \bibinfo{pages}{064026} (\bibinfo{year}{2004}), \eprint{gr-qc/0312030}.

\bibitem[{\citenamefont{Schnetter et~al.}(2004)\citenamefont{Schnetter, Hawley,
  and Hawke}}]{Schnetter-etal-03b}
\bibinfo{author}{\bibfnamefont{E.}~\bibnamefont{Schnetter}},
  \bibinfo{author}{\bibfnamefont{S.~H.} \bibnamefont{Hawley}},
  \bibnamefont{and} \bibinfo{author}{\bibfnamefont{I.}~\bibnamefont{Hawke}},
  \bibinfo{journal}{Class. Quantum Grav.} \textbf{\bibinfo{volume}{21}},
  \bibinfo{pages}{1465} (\bibinfo{year}{2004}), \eprint{gr-qc/0310042}.

\bibitem[{\citenamefont{Cook and Pfeiffer}(2004)}]{Cook:2004kt}
\bibinfo{author}{\bibfnamefont{G.~B.} \bibnamefont{Cook}} \bibnamefont{and}
  \bibinfo{author}{\bibfnamefont{H.~P.} \bibnamefont{Pfeiffer}}
  (\bibinfo{year}{2004}), \bibinfo{note}{gr-qc/0407078}.

\end{thebibliography}


\end{document}